\documentclass[12pt]{article}
\usepackage{amsbsy,amsfonts,amsmath,amssymb,amsthm,ascmac,euscript,enumerate}
\usepackage{graphics,graphicx,color}
\usepackage{url}
\usepackage{hyperref}
\hypersetup{colorlinks,citecolor=blue,linkcolor=blue,urlcolor=black}
\usepackage{bbm}
\newcommand{\indic}{\mathbbm{1}}
\newcommand{\ind}[1]{\indic_{\{#1\}}}

\numberwithin{equation}{section}
\theoremstyle{plain}

\newtheorem{theorem}{Theorem}[section]
\newtheorem{assumption}[theorem]{Assumption}
\newtheorem{corollary}[theorem]{Corollary}
\newtheorem{lemma}[theorem]{Lemma}
\newtheorem{proposition}[theorem]{Proposition}
\newtheorem{remark}[theorem]{Remark}

\newcommand{\cD}{{\cal D}}

\newcommand{\cS}{{\cal S}}

\newcommand{\dc}{d_{\rm c}}

\newcommand{\dpst}{\displaystyle}
\newcommand{\Exp}[1]{\langle#1\rangle}
\newcommand{\lbeq}[1]{\label{eq:#1}}

\newcommand{\mR}{{\mathbb R}}
\newcommand{\mZ}{{\mathbb Z}}
\newcommand{\N}{\mathbb{N}}
\newcommand{\nn}{\nonumber}
\newcommand{\pc}{p_{\rm c}}

\newcommand{\QED}{\hspace*{\fill}\rule{7pt}{7pt}\smallskip}
\newcommand{\Rd}{\mR^d}
\newcommand{\refeq}[1]{(\ref{eq:#1})}
\newcommand{\sss}{\scriptscriptstyle}
\newcommand{\veee}[1]{|\hskip-1.4pt|\hskip-1.4pt|#1|\hskip-1.4pt|\hskip-1.4pt|}
\newcommand{\Zd}{\mZ^d}

\title{Critical two-point function for long-range models with power-law 
couplings: The marginal case for $d\ge\dc$}

\author{
Lung-Chi~Chen\footnote{Department of Mathematical Sciences, 
National Chengchi University, Taiwan}~\footnote{Mathematics Division, 
National Center for Theoretical Sciences, Taiwan},\qquad
Akira~Sakai\footnote{Faculty of Science, Hokkaido University, Japan. 
\url{https://orcid.org/0000-0003-0943-7842}}
}

\begin{document}
\maketitle

\begin{abstract}
Consider the long-range models on $\Zd$ of random walk, self-avoiding walk, 
percolation and the Ising model, whose translation-invariant 1-step 
distribution/coupling coefficient decays as $|x|^{-d-\alpha}$ for some 
$\alpha>0$.  In the previous work \cite{csIV}, we have shown in a unified 
fashion for all $\alpha\ne2$ that, assuming a bound on the ``derivative" of 
the $n$-step distribution (the compound-zeta distribution satisfies 
this assumed bound), the critical two-point function $G_{\pc}(x)$ decays 
as $|x|^{\alpha\wedge2-d}$ above the upper-critical dimension 
$\dc\equiv(\alpha\wedge2)m$, where $m=2$ for self-avoiding walk and the Ising 
model and $m=3$ for percolation. 

In this paper, we show in a much simpler way, 
without assuming a bound on the derivative of the $n$-step distribution, that 
$G_{\pc}(x)$ for the marginal case $\alpha=2$ decays as $|x|^{2-d}/\log|x|$ 
whenever $d\ge\dc$ (with a large spread-out parameter $L$).  This solves 
the conjecture in \cite{csIV}, extended all the way down to $d=\dc$, and 
confirms a part of predictions in physics \cite{bpr14}.  The proof is based on 
the lace expansion and new convolution bounds on power functions with log 
corrections.
\end{abstract}

\tableofcontents

\section{Introduction and the main results}
\subsection{Introduction}\label{ss:intro}
The lace expansion has been successful in rigorously proving mean-field 
critical behavior for various models, such as 
self-avoiding walk \cite{bs85}, percolation \cite{hs90p}, lattice trees and 
lattice animals \cite{hs90l}, oriented percolation \cite{ny93}, the contact 
process \cite{s01}, the classical Ising and $\varphi^4$ models 
\cite{s07,s15}.
It provides (a way to derive) a formal recursion equation for the two-point 
function $G_p(x)$, which is similar to the recursion equation for the 
random-walk Green function $S_p(x)$ generated by the non-degenerate 
(i.e., $D(o)<1$) 1-step distribution $D(x)$ and the fugacity $p\in[0,1]$:
\begin{align}\lbeq{rw-conv}
S_p(x)=\delta_{o,x}+(pD*S_p)(x),
\end{align}
where, and in the rest of the paper, $(f*g)(x)\equiv\sum_yf(y)\,g(x-y)$ 
is the convolution of two functions $f,g$ on $\Zd$.  
The formal recursion equation for $G_p(x)$ is of the form
\begin{align}\lbeq{lace-intro}
G_p(x)=\varPi_p(x)+(\varPi_p*pD*G_p)(x),
\end{align}
where $\varPi_p(x)$ is a series of the model-dependent lace-expansion 
coefficients.  It is natural to expect that, once regularity of $\varPi_p$ (e.g., 
absolute summability) is assured for all $p$ up to the critical point $\pc$, 
the asymptotic behavior of $G_{\pc}(x)$ should be the same (modulo constant 
multiplication) as that for the random-walk Green function $S_1(x)$.  If so, 
then sufficient conditions for the mean-field behavior, called the bubble 
condition for self-avoiding walk and the Ising model \cite{a82,ms93} and the 
triangle condition for percolation \cite{an84}, hold for all dimensions above 
the model-dependent upper-critical dimension $\dc$, which is $2m$ 
for short-range models, where $m=2$ for self-avoiding walk and the Ising model 
and $m=3$ for percolation.

In recent years, long-range models defined by power-law couplings, 
$D(x)\approx|x|^{-d-\alpha}$ for some $\alpha>0$, have attracted more 
attention, due to unconventional critical behavior and crossover phenomena 
(e.g., \cite{apr14,bpr14,csIV,lsw17}).  Under some mild assumptions, 
we have shown \cite[Proposition~2.1]{csIV} that, for $\alpha\ne2$ and 
$d>\alpha\wedge2$, the random-walk Green function $S_1(x)$ is asymptotically 
$\frac{\gamma_\alpha}{v_\alpha}|x|^{\alpha\wedge2-d}$, where
\begin{align}\lbeq{gamma}
\gamma_\alpha=\frac{\Gamma(\frac{d-\alpha\wedge2}2)}{2^{\alpha\wedge2}\pi^{d/2}
 \Gamma(\frac{\alpha\wedge2}2)},&&&&
v_\alpha=\lim_{|k|\to0}\frac{1-\hat D(k)}{|k|^{\alpha\wedge2}}
\equiv\lim_{|k|\to0}\sum_{x\in\Zd}\frac{1-e^{ik\cdot x}}{|k|^{\alpha\wedge2}}D(x).
\end{align}
For short-range models with variance $\sigma^2=\sum_x|x|^2D(x)<\infty$, 
the asymptotic behavior of $S_1(x)$ is well-known to be 
$\frac{d}2\Gamma(\frac{d-2}2)\pi^{-d/2}\sigma^{-2}|x|^{2-d}$, 
which is consistent with \refeq{gamma} for large $\alpha>2$.  The 
crossover occurs at $\alpha=2$, where the variance $\sigma^2$ diverges 
logarithmically and $S_1(x)$ was believed to have a log correction to the 
above standard Newtonian behavior.

An example of $D(x)\approx|x|^{-d-\alpha}$ is the compound-zeta distribution 
(see \refeq{eg123} for the precise definition).  It has been shown \cite{csIV} 
that this long-range distribution for $\alpha\ne2$ also satisfies a certain 
bound on the ``derivative" $|D^{*n}(x)-\frac12(D^{*n}(x+y)+D^{*n}(x-y))|$ of 
the $n$-step distribution.  Thanks to this extra bound, we have shown 
\cite[Theorem~1.2]{csIV} in a unified fashion for all $\alpha\ne2$ that, 
whenever $d>\dc\equiv(\alpha\wedge2)m$ (with a large spread-out parameter 
$L$), there is a model-dependent constant $A$ close to 1 (in fact, $A=1$ for 
$\alpha<2$) such that $G_{\pc}(x)\sim\frac{A}{\pc}S_1(x)$.  One of the key 
elements to showing this result is (slight improvement of) the convolution 
bounds on power functions \cite[Proposition~1.7]{hhs03} that are used to prove 
regularity of $\varPi_p$ in \refeq{lace-intro}.  However, since those convolution 
bounds are not good enough to properly control power functions with log 
corrections, we were unable to achieve an asymptotic result for 
$\alpha=2$, until the current work.

In this paper, we tackle the marginal case $\alpha=2$.  The headlines 
are the following:
\begin{itemize}
\item
$S_1(x)\sim\frac{\gamma_2}{v_2}|x|^{2-d}/\log|x|$ whenever $d>2$, 
where $\gamma_2$ is in \refeq{gamma}, but $v_2$ is redefined as
\begin{align}\lbeq{vdef}
v_2=\lim_{|k|\to0}\frac{1-\hat D(k)}{|k|^2\log(1/|k|)}.
\end{align}
\item
$G_{\pc}(x)\sim\frac1{\pc}S_1(x)$ whenever $d\ge\dc$ (with a large spread-out 
parameter $L$).  This also implies that other critical exponents take on their 
mean-field values for $d\ge\dc$ (including equality).
\end{itemize}
The latter solves the conjecture \cite[(1.29)]{csIV}, extended all the way down 
to $d=\dc$.  It also confirms a part of predictions in physics 
\cite[(3)]{bpr14}: the critical two-point function for percolation was proposed 
to decay as $|x|^{\alpha\wedge(2-\eta)-d}$ whenever $\alpha\ne2-\eta$, where 
$\eta=\eta(d)$ is the anomalous dimension for short-range percolation and is 
believed to be nonzero for $d<6$, and as $|x|^{2-\eta-d}/\log|x|$ whenever 
$\alpha=2-\eta$.

We should emphasize that the proof of the asymptotic result in this paper is 
rather different from the one in \cite{csIV} for $\alpha\ne2$.  In fact, we do 
not require the $n$-step distribution $D^{*n}$ to satisfy the aforementioned 
derivative bound.  Because of this, we can cover a wider class of models to 
which the same result applies, and can simplify the proof to some extent.  
Although the same proof works for $\alpha<2$ (see 
Remark~\ref{remark:new1step} below), we will focus on the marginal case 
$\alpha=2$.

Before closing this subsection, we remark on recent progress in the 
renormalization group analysis for the $O(n)$ model, which is equivalent to 
self-avoiding walk when $n=0$ and to the $n$-component $|\varphi|^4$ model 
when $n\ge1$.  Suppose that the above physics prediction is true for the 
$O(n)$ model as well, and that $\eta>0$ for $d<4$.  Then, we can take a small 
$\varepsilon>0$ to satisfy 
$\alpha=\frac{d+\varepsilon}2\in(\frac{d}2,2-\eta)\ne\varnothing$, hence 
$d=2\alpha-\varepsilon<\dc$, and yet $G_{\pc}(x)$ is proven to decay as 
$|x|^{\alpha-d}$ \cite{lsw17}.  This ``sticking" at the mean-field behavior, 
even below the upper-critical dimension, has been proven by using a rigorous 
version of the $\varepsilon$-expansion.

In the next subsection, we give more precise definitions of the concerned 
models.

\subsection{The models and the main results}
\subsubsection{Random walk}\label{sss:rw}
Let
\begin{align}\lbeq{veee}
\veee{x}_r=\frac\pi2(|x|\vee r)\qquad[x\in\Rd,~1\le r<\infty],
\end{align}
where $|\cdot|$ is the Euclidean norm.  We require the 1-step distribution 
$D(x)$ to be bounded as 
\begin{align}
&D(x)\asymp\tfrac1{L^d}\veee{\tfrac{x}L}_1^{-d-\alpha}\nn\\[5pt]
&\quad\stackrel{\text{def}}\Leftrightarrow\quad\exists c>0,~\forall x\in\Zd,
 ~\forall L\in[1,\infty):~c\le\frac{D(x)}{\frac1{L^d}\veee{\frac{x}L}_1^{-d
 -\alpha}}\le\frac1c,
\end{align}
where $L$ is the spread-out parameter.  

Let $\hat D$ and $D^{*n}$ be the Fourier transform and the $n$-fold 
convolution of $D$, respectively:
\begin{align}
\hat D(k)&=\sum_{x\in\Zd}e^{ik\cdot x}D(x)\qquad[k\in[-\pi,\pi)^d],\\
D^{*n}(x)&=
 \begin{cases}
 \delta_{o,x}&[n=0],\\
 \sum_{y\in\Zd}D^{*(n-1)}(y)\,D(x-y)&[n\ge1].
 \end{cases}
\end{align}
We also require $D$ to satisfy the following properties.

\begin{assumption}[Properties of $\hat D$]\label{assumption:hatD}
There is a $\Delta=\Delta(L)\in(0,1)$ such that
\begin{align}\lbeq{1-hatDbd1}
1-\hat D(k)
 \begin{cases}
 <2-\Delta\quad&[\forall k\in[-\pi,\pi]^d],\\
 >\Delta&[|k|>1/L],
 \end{cases}
\end{align}
while, for $|k|\le1/L$,
\begin{align}\lbeq{1-hatDbd2}
1-\hat D(k)\asymp (L|k|)^{\alpha\wedge2}\times
 \begin{cases}
 1&[\alpha\ne2],\\
 \log\frac\pi{2L|k|}\quad&[\alpha=2].
 \end{cases}
\end{align}
Moreover, there is an $\epsilon>0$ such that, as $|k|\to0$,
\begin{align}\lbeq{1-hatDasy}
1-\hat D(k)=v_\alpha|k|^{\alpha\wedge2}\times
 \begin{cases}
 \big(1+O(L^\epsilon|k|^\epsilon)\big)&[\alpha\ne2],\\
 \big(\log\frac1{L|k|}+O(1)\big)\quad&[\alpha=2],
 \end{cases}
\end{align}
where the constant in the $O(1)$ term is independent of $L$.
\end{assumption}

\begin{assumption}[Bounds on $D^{*n}$]\label{assumption:D}
For $n\in\N$ and $x\in\Zd$,
\begin{gather}
\|D^{*n}\|_\infty\le O(L^{-d})\times
 \begin{cases}
 n^{-d/(\alpha\wedge2)}&[\alpha\ne2],\\
 (n\log\frac{\pi n}2)^{-d/2}\quad&[\alpha=2],
 \end{cases}\lbeq{Dnsupbd}\\
D^{*n}(x)\le n\frac{O(L^{\alpha\wedge2})}{\veee{x}_L^{d+\alpha\wedge2}}\times
 \begin{cases}
 1&[\alpha\ne2],\\
 \log\veee{\frac{x}L}_1\quad&[\alpha=2].
 \end{cases}\lbeq{Dnxbd}
\end{gather}
\end{assumption}

It has been shown \cite{csI,csIII,csIV} that the following $D$ is one of the 
examples that satisfy all the properties in the above assumptions:
\begin{align}\lbeq{eg0}
D(x)=
 \begin{cases}
 \dpst\frac{\veee{x}_L^{-d-\alpha}}{\sum_{y\in\Zd\setminus\{o\}}\veee{y}_L^{-d
  -\alpha}}\quad&[x\ne o],\\[1pc]
 0&[x=o].
 \end{cases}
\end{align}
Another such example is the following compound-zeta distribution \cite{csIV}:
\begin{align}\lbeq{eg123}
D(x)=\sum_{t\in\N}U_L^{*t}(x)\,T_\alpha(t)\qquad[x\in\Zd],
\end{align}
where, with a probability distribution $h$ on $[-1,1]^d\subset\Rd$ and the 
Riemann-zeta function $\zeta(s)=\sum_{t\in\N}t^{-s}$, 
\begin{align}
U_L(x)&=\frac{h(x/L)}{\sum_{y\in\Zd\setminus\{o\}}h(y/L)}\qquad[x\in\Zd],\\
T_\alpha(t)&=\frac{t^{-1-\alpha/2}}{\zeta(1+\alpha/2)}\qquad[t\in\N].
\end{align}
We assume that the distribution $h$ is bounded, non-degenerate, 
$\Zd$-symmetric and piecewise continuous, such as 
$h(x)=2^{-d}\ind{\|x\|_\infty\le1}$.  

Since the proof of 
\refeq{Dnsupbd} for $\alpha=2$ is only briefly explained in \cite[(1.19)]{csIV}, 
we will provide a full proof in Section~\ref{s:rw}.

Let $S_p$ be the random-walk Green function generated by the 1-step 
distribution $D$:
\begin{align}\lbeq{green}
S_p(x)=\sum_{\omega:o\to x}p^{|\omega|}\prod_{j=1}^{|\omega|}
 D(\omega_j-\omega_{j-1})\qquad[x\in\Zd],
\end{align}
where $o\in\Zd$ is the origin, $p\ge0$ is the fugacity and $|\omega|$ is the 
length of a path $\omega=(\omega_0,\omega_1,\dots,\omega_{|\omega|})$.  
By convention, the contribution from the zero-step walk is the Kronecker delta 
$\delta_{o,x}$.  It is convergent as long as $p<1$ or $p=1$ with 
$d>\alpha\wedge2$.  One of the main results of this paper is completion of 
the asymptotic picture of $S_1$ for all $\alpha>0$, as follows.

\begin{theorem}\label{theorem:S}
Let $d>\alpha\wedge2$ and suppose $D$ satisfies 
Assumptions~\ref{assumption:hatD}--\ref{assumption:D}.  
Then, for any $p\in[0,1]$,
\begin{align}\lbeq{Squbd}
S_p(x)-\delta_{o,x}\le\frac{O(L^{-\alpha\wedge2})}{\veee{x}_L^{d-\alpha
 \wedge2}}\times
 \begin{cases}
 1&[\alpha\ne2],\\
 \dfrac1{\log\veee{\frac{x}L}_1}\quad&[\alpha=2].
 \end{cases}
\end{align}
Moreover, there are $\epsilon,\eta>0$ such that, for $L^{1+\eta}<|x|\to\infty$,
\begin{align}\lbeq{S1asy}
S_1(x)=\frac{\gamma_\alpha/v_\alpha}{|x|^{d-\alpha\wedge2}}\times
 \begin{cases}
 \dpst\bigg(1+\frac{O(L^{\epsilon})}{|x|^\epsilon}\bigg)&[\alpha\ne2],\\[1pc]
 \dpst\frac1{\log|x|}\bigg(1+\frac{O(1)}{(\log|x|)^\epsilon}\bigg)\quad
  &[\alpha=2],
 \end{cases}
\end{align}
where the constant in the $O(1)$ term is independent of $L$.
\end{theorem}

\subsubsection{Self-avoiding walk}
Self-avoiding walk (sometimes abbreviated as SAW) is a model for linear 
polymers.  Taking into account the exclusion-volume effect among constituent 
monomers, we define the SAW two-point function as
\begin{align}\lbeq{SAW-2pt}
G_p(x)=\sum_{\omega:o\to x}p^{|\omega|}\prod_{j=1}^{|\omega|}
 D(\omega_j-\omega_{j-1})\prod_{s<t}(1-\delta_{\omega_s,\omega_t}),
\end{align}
where the contribution from the zero-step walk is $\delta_{o,x}$, just as in 
\refeq{green}.  Notice that the difference between \refeq{green} and 
\refeq{SAW-2pt} is the last product, which is either 0 or 1 depending on 
whether $\omega$ intersects itself or does not.  Because of this suppressing 
factor, the sum called the susceptibility 
\begin{align}\lbeq{susceptibility}
\chi_p=\sum_{x\in\Zd}G_p(x)
\end{align}
is not bigger than $\sum_{x\in\Zd}S_p(x)$, which is $(1-p)^{-1}$ when $p$ is 
smaller than the radius of convergence 1, and therefore the critical point 
\begin{align}\lbeq{critpt}
\pc=\sup\{p:\chi_p<\infty\}
\end{align}
must be at least 1.  It is known \cite{ms93} that, if the bubble condition 
\begin{align}\lbeq{bubblecond}
G_{\pc}^{*2}(o)=\sum_{x\in\Zd}G_{\pc}(x)^2<\infty
\end{align}
holds, then 
\begin{align}\lbeq{chiMF}
\chi_p\asymp(\pc-p)^{-1},
\end{align}
meaning that the critical exponent for $\chi_p$ takes on its mean-field value 1.

\subsubsection{Percolation}
Percolation is a model for random media.  Each bond $\{u,v\}\subset\Zd$ is 
assigned to be either occupied or vacant, independently of the other bonds.  
The probability of a bond $\{u,v\}$ being occupied is defined as $pD(v-u)$,
where $p\ge0$ is the percolation parameter.  Since $D$ is a probability 
distribution, the expected number of occupied bonds per vertex equals 
$p\sum_{x\ne o}D(x)=p(1-D(o))$.  Let $G_p(x)$ denote the percolation 
two-point function, which is the probability that there is a self-avoiding path 
of occupied bonds from $o$ to $x$.  By convention, $G_p(o)=1$.

For percolation, the susceptibility $\chi_p$ in \refeq{susceptibility} equals 
the expected number of vertices connected from $o$.  It is known 
\cite{an84} that there is a critical point $\pc$ defined as in \refeq{critpt} 
such that $\chi_p$ is finite if and only if $p<\pc$ and diverges as 
$p\uparrow\pc$.  It is also known that, if the triangle condition
\begin{align}\lbeq{trianglecond}
G_{\pc}^{*3}(o)=\sum_{x\in\Zd}G_{\pc}(x)\,G_{\pc}^{*2}(x)<\infty
\end{align}
holds, then $\chi_p$ diverges in the same way as \refeq{chiMF}.  

There is another order parameter $\theta_p$ called the percolation probability, 
which is the probability of the origin $o$ being connected to infinity.  It is 
known \cite{ab87,dct16,m86} that $\pc$ in \refeq{critpt} can be characterized 
as $\inf\{p\ge0:\theta_p>0\}$ and that, if the triangle condition 
\refeq{trianglecond} holds, then
\begin{align}\lbeq{thetaMF-perc}
\theta_p\asymp p-\pc,
\end{align}
meaning that the critical exponent for $\theta_p$ takes on 
its mean-field value 1, i.e., the value for the survival probability of the 
branching process.

\subsubsection{The Ising model}
The Ising model is a model for magnets.  Let $\Lambda\subset\Zd$ and define 
the Hamiltonian (under the free-boundary condition) for a spin configuration 
$\varphi=\{\varphi_v\}_{v\in\Lambda}\in\{\pm1\}^\Lambda$ as
\begin{align}
H_\Lambda(\varphi)=-\sum_{\{u,v\}\subset\Lambda}J_{u,v}\varphi_u
 \varphi_v,
\end{align}
where $J_{u,v}=J_{o,v-u}\ge0$ is the ferromagnetic coupling and is to satisfy 
the relation
\begin{align}
D(x)=\frac{\tanh(\beta J_{o,x})}{\sum_{y\in\Zd}\tanh(\beta J_{o,y})},
\end{align}
where $\beta\ge0$ is the inverse temperature.  Let
\begin{align}\lbeq{therm-av}
\Exp{\varphi_o\varphi_x}_{\beta,\Lambda}=\sum_{\varphi\in\{\pm1\}^\Lambda}
 \varphi_o\varphi_x\;e^{-\beta H_\Lambda(\varphi)}\Bigg/\sum_{\varphi\in\{\pm
 1\}^\Lambda}e^{-\beta H_\Lambda(\varphi)}.
\end{align}
Using $p=\sum_{x\in\Zd}\tanh(\beta J_{o,x})$, we define the Ising
two-point function $G_p(x)$ as a unique infinite-volume limit of 
$\Exp{\varphi_o\varphi_x}_{\beta,\Lambda}$:
\begin{align}
G_p(x)=\lim_{\Lambda\uparrow\Zd}\Exp{\varphi_o\varphi_x}_{\beta,\Lambda}.
\end{align}

It is known \cite{l74} that the susceptibility $\chi_p$ defined as in 
\refeq{susceptibility} is finite if and only if $p<\pc$ and diverges as 
$p\uparrow\pc$.  It is also known \cite{abf87,dct16} that $\pc$ 
is unique in the sense that the spontaneous magnetization 
\begin{align}
\theta_p=\sqrt{\lim_{|x|\to\infty}G_p(x)}
\end{align}
also exhibits a phase transition at $\pc$.  (Unlike the case for percolation, 
the continuity of $\theta_p$ in $p$ has been proven for all dimensions, as 
long as $J_{o,x}$ satisfies a strong symmetry condition called the reflection 
positivity \cite{adcs15}.)  Furthermore, it is known \cite{a82,af86} that, if the 
bubble condition \refeq{bubblecond} holds for the critical Ising model, then
\begin{align}\lbeq{thetaMF-Ising}
\chi_p\asymp(\pc-p)^{-1},&&
\theta_p\asymp\sqrt{p-\pc},
\end{align}
meaning that the critical exponents for $\chi_p$ and $\theta_p$ take on their 
mean-field values 1 and $1/2$, respectively.

\subsubsection{The main results}
Let
\begin{align}\lbeq{dcdef}
\dc=(\alpha\wedge2)\times m,&&
m=\begin{cases}
 2\quad&[\text{SAW and Ising}],\\
 3&[\text{percolation}],
 \end{cases}
\end{align}
where $m$ is the number of $G_{\pc}$ involved in the bubble/triangle 
conditions \refeq{bubblecond} and \refeq{trianglecond}.  

In the previous paper \cite{csIV}, we investigated asymptotic behavior of  
$G_{\pc}(x)$ for $\alpha\ne2$, $d>\dc$ and $L\gg1$ (see 
Theorem~\ref{theorem:previous}).  In the current paper, we investigate 
the marginal case $\alpha=2$, for which the variance of $D$ diverges 
logarithmically, and prove the following:

\begin{theorem}\label{theorem:main}
Let $\alpha=2$ and $d\ge\dc$ (including equality) and suppose that $D$ 
satisfies Assumptions~\ref{assumption:hatD}--\ref{assumption:D}.  
Then there is a model-dependent $L_0<\infty$ such that, for any $L\ge L_0$,
\begin{align}\lbeq{IRbd}
G_{\pc}(x)\le\delta_{o,x}+\frac{O(L^{-2})}{\veee{x}_L^{d-2}\log\veee{\frac{x}
 L}_1}.
\end{align}
Moreover, there is an $\epsilon>0$ such that, as $|x|\to\infty$,
\begin{align}\lbeq{main}
G_{\pc}(x)=\frac1{\pc}\frac{\gamma_2/v_2}{|x|^{d-2}\log|x|}\bigg(1+\frac{O(1)}
 {(\log|x|)^\epsilon}\bigg),
\end{align}
where the $O(1)$ term is independent of $L$.
\end{theorem}

Due to the log correction to the standard Newtonian behavior in 
\refeq{IRbd}--\refeq{main}, we can show that the bubble/triangle conditions 
hold, even at the critical dimension $d=\dc$.  For example, the tail of the sum 
in the bubble condition \refeq{bubblecond} can be estimated, for any 
$R>1$, as 
\begin{align}
\sum_{x:|x|>R}G_{\pc}(x)^2\approx\int_R^\infty\frac{\mathrm{d}r}r~
 \frac{r^{4-d}}{(\log r)^2},
\end{align}
which is finite even when $d=4$, due to the log-squared term in the 
denominator.  Also, by the convolution bounds in Lemma~\ref{lemma:conv-bds} 
below, 
which is one of the novelties of this paper, we can show that $G_{\pc}^{*2}(x)$ 
for $d\ge4$ is bounded above by a multiple of $|x|^{4-d}/\log|x|$.  Therefore, 
the tail of the sum in the triangle condition \refeq{trianglecond} can be 
estimated as
\begin{align}
\sum_{x:|x|>R}G_{\pc}(x)\,G_{\pc}^{*2}(x)\approx\int_R^\infty
 \frac{\mathrm{d}r}r~\frac{r^{6-d}}{(\log r)^2},
\end{align}
which is finite even when $d=6$, again due to the log-squared term in the 
denominator.  Therefore:

\begin{corollary}
The mean-field results \refeq{chiMF}, \refeq{thetaMF-perc} and 
\refeq{thetaMF-Ising} hold for all three models with $\alpha=2$ and 
sufficiently large $L$, in dimensions $d\ge\dc$ (including equality).
\end{corollary}

\smallskip

\begin{remark}
{\rm
\begin{enumerate}
\item
In the previous paper \cite{csIV}, we investigated the other case 
$\alpha\ne2$ and proved the following:

\begin{theorem}[Theorems~1.2 and 3.3 of \cite{csIV}]\label{theorem:previous}
Let $\alpha\ne2$ and $d>\dc$ and suppose that $D$ satisfies 
Assumptions~\ref{assumption:hatD}--\ref{assumption:D} and the following 
bound on the ``derivative" of $D^{*n}$: for $n\in\N$ and $x,y\in\Zd$ with 
$|y|\le\frac13|x|$,
\begin{align}\lbeq{Dnxdiffbd}
\bigg|D^{*n}(x)-\frac{D^{*n}(x+y)+D^{*n}(x-y)}2\bigg|\le n\,\frac{O(L^{\alpha
 \wedge2})\,\veee{y}_L^2}{\veee{x}_L^{d+\alpha\wedge2+2}}.
\end{align}
Then, there is a model-dependent $L_0<\infty$ such that, 
for any $L\ge L_0$,
\begin{align}\lbeq{IRbdprevious}
G_{\pc}(x)\le\delta_{o,x}+\frac{O(L^{-\alpha\wedge2})}{\veee{x}_L^{d-\alpha
 \wedge2}}.
\end{align}
As a result, the bubble/triangle conditions \refeq{bubblecond} and 
\refeq{trianglecond} hold, and therefore the critical exponents for $\chi_p$ 
and $\theta_p$ take on their respective mean-field values.  Moreover, 
there are $A=1+O(L^{-2})\ind{\alpha>2}$ and $\epsilon>0$ such that, as 
$|x|\to\infty$,
\begin{align}\lbeq{previous}
G_{\pc}(x)=\frac{A}{\pc}\frac{\gamma_\alpha/v_\alpha}{|x|^{d-\alpha\wedge2}}
 \bigg(1+\frac{O(L^\epsilon)}{|x|^\epsilon}\bigg).
\end{align}
\end{theorem}

\smallskip

The extra assumption \refeq{Dnxdiffbd} is hard to verify in a general setup.  
However, we have shown \cite{csIV} that the compound-zeta distribution 
\refeq{eg123} for $\alpha\ne2$ satisfies \refeq{Dnxdiffbd}.  In fact, as 
explained in Section~\ref{ss:xIRbd} (see also Remark~\ref{remark:new1step}), 
the proof of Theorem~\ref{theorem:main} for $\alpha=2$ also works for the 
case $\alpha<2$, so that we do not have to require \refeq{Dnxdiffbd} 
for $\alpha\le2$, but not for $\alpha>2$.  This is somewhat related 
to the fact that the 
multiplicative constant $A$ in \refeq{previous} becomes 1 for $\alpha\le2$.
\item
The possibility to extend the mean-field results down to $d=\dc$ was already 
hinted in \cite[Theorem~1.1]{hhs08}, where we have shown that, for $d>\dc$ 
and $L\gg1$, the Fourier transform $\hat G_p(k)$ obeys the following infrared 
bound, uniformly in $k\in[-\pi,\pi]^d$ and $p<\pc$:
\begin{align}\lbeq{kIRbd}
\hat G_p(k)=\frac{1+O(\delta_m)}{\chi_p^{-1}+p(1-\hat D(k))},
\end{align}
where
\begin{align}
\delta_m=\int_{[-\pi,\pi]^d}\frac{\mathrm{d}^dk}{(2\pi)^d}~\frac{\hat D(k)^2}
 {(1-\hat D(k))^m}.
\end{align}
In fact, we can follow the same line of proof of \cite[Theorem~1.1]{hhs08} to 
obtain \refeq{kIRbd}, as long as $\delta_m$ is sufficiently small.  However, 
for $\alpha=2$ and $d\ge\dc$ (including equality), we have
\begin{align}
\delta_m\le\underbrace{\int_{|k|>1/L}\frac{\mathrm{d}^dk}{(2\pi)^d}~
 \frac{\hat D(k)^2}{\Delta^m}}_{\because\,\refeq{1-hatDbd1}}
 +\underbrace{O(L^{-2m})\int_{|k|\le1/L}\frac{\mathrm{d}^dk}{(|k|^2
 \log\frac\pi{2L|k|})^m}}_{\because\,\refeq{1-hatDbd2}}=O(L^{-d}).
\end{align}
Therefore, by taking $L$ sufficiently large and using monotonicity in $p$,
we obtain
\begin{align}
G_{\pc}^{*m}(o)=\lim_{p\uparrow\pc}G_p^{*m}(o)=\lim_{p\uparrow\pc}
 \int_{[-\pi,\pi]^d}\frac{\mathrm{d}^dk}{(2\pi)^d}~\hat G_p(k)^m<\infty,
\end{align}
as long as $d\ge\dc$, hence the mean-field results for all $d\ge\dc$.
\end{enumerate}
}
\end{remark}

\section{Analysis for the underlying random walk}\label{s:rw}
In Section~\ref{ss:green}, we prove Theorem~\ref{theorem:S} for $\alpha=2$ 
(the results for $\alpha\ne2$ have been proven in \cite{csIV}).  
In Section~\ref{ss:D*n}, we complete the proof of \refeq{Dnsupbd} for $\alpha=2$.

\subsection{Proof of Theorem~\ref{theorem:S}}\label{ss:green}
The results for $\alpha\ne2$ are already proven in 
\cite[Proposition~2.1]{csIV}.  The proof of \refeq{Squbd} for $\alpha=2$ is 
easy, as we split the sum at 
$N\equiv\veee{\frac{x}L}_1^2/\log\veee{\frac{x}L}_1$ and use \refeq{Dnxbd} for 
$n\le N$ and \refeq{Dnsupbd} for $n\ge N$, as follows:
\begin{align}\lbeq{Spubd-proof}
S_q(x)-\delta_{o,x}&\le\sum_{n=1}^ND^{*n}(x)+\sum_{n=N}^\infty\|D^{*n}
 \|_\infty\nn\\
&\le O(L^{-d})\bigg(\frac{\log\veee{\frac{x}L}_1}{\veee{\frac{x}L}_1^{d+2}}
 \sum_{n=1}^Nn+\sum_{n=N}^\infty(n\log n)^{-d/2}\bigg)\nn\\
&\le O(L^{-d})\bigg(\frac{\log\veee{\frac{x}L}_1}{\veee{\frac{x}L}_1^{d+2}}
 N^2+\frac{N^{1-d/2}}{(\log N)^{d/2}}\bigg)~
 =\frac{O(L^{-d})\veee{\frac{x}L}_1^{2-d}}{\log\veee{\frac{x}L}_1}.
\end{align}

It remains to show \refeq{S1asy} for $\alpha=2$.  First, we rewrite $S_1(x)$ 
for $d>2$ as
\begin{align}
S_1(x)=\int_{[-\pi,\pi]^d}\frac{\text{d}^dk}{(2\pi)^d}\,\frac{e^{-ik\cdot x}}
 {1-\hat D(k)}&=\int_0^\infty\text{d}t\int_{[-\pi,\pi]^d}\frac{\text{d}^dk}
 {(2\pi)^d}\,e^{-ik\cdot x-t(1-\hat D(k))}.
\end{align}
Let 
\begin{align}\lbeq{Tdef}
\mu\in(0,\tfrac2{d+2}),\hskip4pc
T=\frac{(\frac{|x|}L)^2}{(\log\frac{|x|}L)^{1+\mu}}.
\end{align}
Then, for $|x|>L^{1+\eta}$ (so that $\delta_{o,x}=0$),
\begin{eqnarray}\lbeq{I1ubd}
I_1&\equiv&\int_0^T\text{d}t\int_{[-\pi,\pi]^d}\frac{\text{d}^dk}{(2\pi)^d}\,
 e^{-ik\cdot x-t(1-\hat D(k))}\text{d}t\nn\\
&=&\int_0^T\text{d}t\,e^{-t}\sum_{n=0}^\infty\frac{t^n}{n!}\int_{[-\pi,\pi]^d}
 \frac{\text{d}^dk}{(2\pi)^d}\,e^{-ik\cdot x}\hat D(k)^n\nn\\
&=&\int_0^T\text{d}t\,e^{-t}\bigg(\delta_{o,x}+\sum_{n=1}^\infty\frac{t^n}{n!}
 D^{*n}(x)\bigg)\nn\\
&\stackrel{\refeq{Dnxbd}}\le&\frac{O(L^2)\log\frac{|x|}L}{|x|^{d+2}}\,T^2~
 =\frac{O(L^{-2})|x|^{2-d}}{(\log\frac{|x|}L)^{1+2\mu}},
\end{eqnarray}
which is an error term.

Next, we investigate $S_1(x)-I_1$.  Let 
\begin{align}\lbeq{Rdef}
\omega=\frac1{\eta\log L}\in(0,1),\hskip4pc
LR=\bigg(\frac{|x|}L\bigg)^{-\omega}.
\end{align}
Then, we can rewrite $S_1(x)-I_1$ as
\begin{align}\lbeq{S1dec2}
S_1(x)-I_1&=\int_T^\infty\text{d}t\int_{[-\pi,\pi]^d}\frac{\text{d}^dk}{(2\pi)^d}
 \,e^{-ik\cdot x-t(1-\hat D(k))}\text{d}t\nn\\
&=\int_T^\infty\text{d}t\int_{|k|\le R}\frac{\text{d}^d
 k}{(2\pi)^d}e^{-ik\cdot x-v_2t|k|^2\log\frac1{L|k|}}+\sum_{j=2}^4I_j,
\end{align}
where
\begin{align}
I_2&=\int_T^\infty\text{d}t\int_{|k|\le R}\frac{\text{d}^dk}{(2\pi)^d}\,
 e^{-ik\cdot x}\Big(e^{-t(1-\hat D(k))}-e^{-v_2t|k|^2\log\frac1{L|k|}}\Big),\\
I_3&=\int_{R<|k|\le1/L}\frac{\text{d}^dk}{(2\pi)^d}\,\frac{e^{-ik\cdot x-T
 (1-\hat D(k))}}{1-\hat D(k)},\\
I_4&=\int_{[-\pi,\pi]^d}\frac{\text{d}^dk}{(2\pi)^d}\,\frac{e^{-ik\cdot x-T
 (1-\hat D(k))}}{1-\hat D(k)}\,\ind{|k|>1/L}.
\end{align}

For $I_2$, we first note that, by \refeq{1-hatDasy}, 
\begin{align}
\Big|e^{-t(1-\hat D(k))}-e^{-v_2t|k|^2\log\frac1{L|k|}}\Big|\le O(L^2)t|k|^2
 e^{-v_2t|k|^2\log\frac1{L|k|}}.
\end{align}
Let $s=v_2t|k|^2\log\frac1{L|k|}$ and $r=|k|\le R$.  Since $|x|>L^{1+\eta}$, 
we have
\begin{gather}
\frac{\text{d}s}s=\bigg(2-\frac1{\log\frac1{Lr}}\bigg)\frac{\text{d}r}r\ge\bigg(
 2-\frac1{\log\frac1{LR}}\bigg)\frac{\text{d}r}r>\bigg(2-\frac1{\omega\eta
 \log L}\bigg)\frac{\text{d}r}r=\frac{\text{d}r}r.
\end{gather}
Therefore, for $d>2$,
\begin{eqnarray}\lbeq{I2ubd}
|I_2|&\le&O(L^2)\int_T^\infty\text{d}t~t\int_0^R\frac{\text{d}r}r~r^{d+2}
 e^{-v_2tr^2\log\frac1{Lr}}\nn\\
&\le&O(L^2)\int_T^\infty\text{d}t~t\int_0^{v_2tR^2\log\frac1{LR}}\frac{\text{d}s}
 s~\bigg(\frac{s}{v_2t\log\frac1{LR}}\bigg)^{(d+2)/2}e^{-s}\nn\\
&\stackrel{\refeq{Rdef}}\le&O(L^{-d})\bigg(\log\frac{|x|}L\bigg)^{-(d+2)/2}
 T^{1-d/2}\nn\\
&\stackrel{\refeq{Tdef}}=&\frac{O(L^{-2})|x|^{2-d}}{(\log\frac{|x|}L)^{2-(d-2)
 \mu/2}},
\end{eqnarray}
which is an error term because
\begin{align}
2-\frac{(d-2)\mu}2\stackrel{\refeq{Tdef}}>2-\frac{d-2}{d+2}=1+\frac4{d+2}>1.
\end{align}

For $I_3$, since \refeq{1-hatDbd2} holds and 
$\log\frac\pi{2L|k|}\ge\log\frac\pi2>0$ for $|k|\le1/L$, 
there is a $c>0$ such that 
\begin{align}
|I_3|\le O(L^{-2})\int_R^{1/L}\frac{\text{d}r}r~r^{d-2}e^{-cL^2Tr^2}=O(L^{-d})
 T^{1-d/2}\int_{cL^2TR^2}^{cT}\frac{\text{d}s}s~s^{(d-2)/2}e^{-s}.
\end{align}
Since $TR^2\to\infty$ as $|x|\to\infty$ (cf., \refeq{Tdef} and \refeq{Rdef}), 
the integral is bounded by a multiple of  
$(L^2TR^2)^{(d-4)/2}e^{-cL^2TR^2}$, which is a bound on the incomplete gamma 
function.  Therefore, for $N\in\N$ large enough to 
ensure $2N+4>d$, 
\begin{align}\lbeq{I3ubd}
|I_3|\le O(L^{-d})\frac{(LR)^{d-4}}Te^{-cL^2TR^2}
&\le\frac{(LR)^{d-4}}T\frac{O(L^{-d})}{(L^2TR^2)^N}\nn\\
&=\frac{O(L^{-2+(2N+4-d)(1-\omega)})(\log\frac{|x|}L)^{(1+\mu)(N+1)}}
 {|x|^{d-2+(2N+4-d)(1-\omega)}}\nn\\
&\hskip-1.1pc\stackrel{|x|>L^{1+\eta}}\le\frac{O(L^{-2})(\log\frac{|x|}
 L)^{(1+\mu)(N+1)}}{|x|^{d-2+(2N+4-d)(1-\omega)\eta/(1+\eta)}},
\end{align}
which is an error term.

For $I_4$, we use \refeq{1-hatDbd1} and a similar argument to \refeq{I3ubd} 
to obtain that, for $N\in\N$ large enough to ensure 
$2(N\eta-1)/(1+\eta)>d-2$,
\begin{align}\lbeq{I4ubd}
|I_4|\le O(1)e^{-T\Delta}
&\le\frac{O(1)}{T^N}\le\frac{O(L^{2N})(\log\frac{|x|}L)^N}{|x|^{2N}}
 \stackrel{|x|>L^{1+\eta}}\le\frac{O(L^{-2})(\log\frac{|x|}L)^N}
 {|x|^{2(N\eta-1)/(1+\eta)}},
\end{align}
which is an error term.

So far, we have obtained
\begin{align}\lbeq{S1dec2rewr}
S_1(x)=\int_T^\infty\text{d}t\int_{|k|\le R}\frac{\text{d}^dk}{(2\pi)^d}e^{-ik
 \cdot x-v_2t|k|^2\log\frac1{L|k|}}+\sum_{j=1}^4I_j.
\end{align}
To investigate the above integral, we introduce $\xi\equiv x/|x|$ and change 
variables as $\kappa=|x|k$.  
Then, by changing time variables as $\tau=\frac{v_2t}{|x|^2}\log\frac{|x|}L$, 
the integral in \refeq{S1dec2} can be written as
\begin{align}\lbeq{S1dec3}
&|x|^{-d}\int_T^\infty\text{d}t\int_{|\kappa|\le|x|R}\frac{\text{d}^d\kappa}
 {(2\pi)^d}~\exp\bigg(-i\kappa\cdot\xi-\frac{v_2t|\kappa|^2}{|x|^2}
 \log\frac{|x|}{L|\kappa|}\bigg)\nn\\
&=\frac{|x|^{2-d}}{v_2\log\frac{|x|}L}\int_{\frac{v_2T}{|x|^2}\log\frac{|x|}
 L}^\infty\text{d}\tau\int_{|\kappa|\le|x|R}\frac{\text{d}^d\kappa}
 {(2\pi)^d}~\exp\Bigg(-i\kappa\cdot\xi-\tau|\kappa|^2\frac{\log
 \frac{|x|}{L|\kappa|}}{\log\frac{|x|}L}\Bigg)\nn\\
&=\frac{|x|^{2-d}}{v_2\log\frac{|x|}L}\bigg(\int_0^\infty\text{d}\tau\int_{\Rd}
 \frac{\text{d}^d\kappa}{(2\pi)^d}~e^{-i\kappa\cdot\xi-\tau|\kappa|^2}
 -\sum_{j=1}^3M_j\bigg),
\end{align}
where
\begin{align}
M_1&=\int_0^{\frac{v_2T}{|x|^2}\log\frac{|x|}L}\text{d}\tau\int_{\Rd}
 \frac{\text{d}^d\kappa}{(2\pi)^d}~e^{-i\kappa\cdot\xi-\tau|\kappa|^2},\\
M_2&=\int_{\frac{v_2T}{|x|^2}\log\frac{|x|}L}^\infty\text{d}\tau\int_{|\kappa|
 >|x|R}\frac{\text{d}^d\kappa}{(2\pi)^d}~e^{-i\kappa\cdot\xi-\tau|\kappa|^2},\\
M_3&=\int_{\frac{v_2T}{|x|^2}\log\frac{|x|}L}^\infty\text{d}\tau\int_{|\kappa|
 \le|x|R}\frac{\text{d}^d\kappa}{(2\pi)^d}~e^{-i\kappa\cdot\xi-\tau
 |\kappa|^2}\Bigg(1-\exp\bigg(\tau|\kappa|^2\frac{\log|\kappa|}
 {\log\frac{|x|}L}\bigg)\Bigg).
\end{align}
Notice that the first term in the parentheses in \refeq{S1dec3} gives the 
leading term:
\begin{align}\lbeq{S1dec4}
\int_0^\infty\text{d}\tau\int_{\Rd}\frac{\text{d}^d\kappa}{(2\pi)^d}~e^{-i
 \kappa\cdot\xi-\tau|\kappa|^2}=\int_0^\infty\text{d}\tau~\frac{e^{-1/(4
 \tau)}}{(4\pi\tau)^{d/2}}&=\frac{\Gamma(\frac{d-2}2)}{4\pi^{d/2}}=\gamma.
\end{align}

For $M_1$, since 
$|x|^2/(v_2T\log\frac{|x|}L)=\frac{L^2}{v_2}(\log\frac{|x|}L)^\mu\to\infty$, 
we obtain that, for $N\in\N$ large enough to ensure $2N+4>d$, 
\begin{align}
M_1=\int_0^{\frac{v_2T}{|x|^2}\log\frac{|x|}L}\text{d}\tau~\frac{e^{-1/(4\tau)}}
 {(4\pi\tau)^{d/2}}
&=\frac1{4\pi^{d/2}}\int_{|x|^2/(4v_2T\log\frac{|x|}L)}^\infty\frac{\text{d}s}s~
 s^{(d-2)/2}e^{-s}\nn\\
&\le O(1)~\bigg(\frac{|x|^2}{v_2T\log\frac{|x|}L}\bigg)^{(d-4)/2}e^{-|x|^2
 /(4v_2T\log\frac{|x|}L)}.
\end{align}
Using the exponentially decaying term yields
\begin{align}\lbeq{M1bd}
M_1\stackrel{\forall N}\le\frac{O(1)}{(\log\frac{|x|}L)^{(2N+4-d)\mu/2}},
\end{align}
which gives an error term as long as $2N+4>d$.

For $M_2$, changing the order of integrations and changing variables as 
$r=|\kappa|^2\frac{v_2T}{|x|^2}\log\frac{|x|}L$ yields
\begin{align}
|M_2|&\le\int_{|\kappa|>|x|R}\frac{\text{d}^d\kappa}{(2\pi)^d}\int_{\frac{v_2T}
 {|x|^2}\log\frac{|x|}L}^\infty\text{d}\tau~e^{-\tau|\kappa|^2}\nn\\
&=\int_{|\kappa|>|x|R}\frac{\text{d}^d\kappa}{(2\pi)^d}~\frac1{|\kappa|^2}
 \exp\bigg(-|\kappa|^2\frac{v_2T}{|x|^2}\log\frac{|x|}L\bigg)\nn\\
&=O(1)\,\bigg(\frac{|x|^2}{v_2T\log\frac{|x|}L}\bigg)^{(d-2)/2}
\int_{v_2TR^2\log\frac{|x|}L}^\infty\frac{\text{d}r}r~r^{(d-2)/2}e^{-r}\nn\\
&=O(1)\,(|x|R)^{d-4}\frac{|x|^2}{v_2T\log\frac{|x|}L}\,e^{-v_2TR^2\log\frac{|x|}L}.
\end{align}
Using the exponentially decaying term and $|x|>L^{1+\eta}$ as in  
\refeq{I3ubd}--\refeq{I4ubd}, we obtain that, for $N\in\N$ large enough 
to ensure $2N+4>d$, 
\begin{align}\lbeq{M2bd}
|M_2|\le\frac{O(1)(\log\frac{|x|}L)^{(N+1)\mu}}{(\frac{|x|}L)^{(2N+4-d)(1
 -\omega)}}\stackrel{|x|>L^{1+\eta}}\le\frac{O(1)(\log\frac{|x|}L)^{(N+1)
 \mu}}{|x|^{(2N+4-d)(1-\omega)\eta/(1+\eta)}},
\end{align}
which gives another error term.

For $M_3$, we first note that, since $|\kappa|\le|x|R=(|x|/L)^{1-\omega}$, 
\begin{align}
\Bigg|1-\exp\bigg(\tau|\kappa|^2\frac{\log|\kappa|}{\log\frac{|x|}L}\bigg)
 \Bigg|&\le\tau|\kappa|^2\frac{|\log|\kappa||}{\log\frac{|x|}L}\,\exp\bigg(
 \tau|\kappa|^2\frac{\log|x|R}{\log\frac{|x|}L}\bigg)\nn\\
&=\tau|\kappa|^2\frac{|\log|\kappa||}{\log\frac{|x|}L}\,e^{(1-\omega)
 \tau|\kappa|^2}.
\end{align}
Then, by changing the order of integrations and changing variables as 
$s=\omega|\kappa|^2\frac{v_2T}{|x|^2}\log\frac{|x|}L$, we obtain
\begin{align}
|M_3|&\le\frac1{\log\frac{|x|}L}\int_{|\kappa|\le|x|R}\frac{\text{d}^d\kappa}
 {(2\pi)^d}~|\kappa|^2|\log|\kappa||\int_{\frac{v_2T}{|x|^2}\log\frac{|x|}
 L}^\infty\text{d}\tau~\tau\,e^{-\omega\tau|\kappa|^2}\nn\\
&=\frac1{\log\frac{|x|}L}\int_{|\kappa|\le|x|R}\frac{\text{d}^d\kappa}{(2\pi)^d}
 ~\frac{|\log|\kappa||}{\omega^2|\kappa|^2}\bigg(1+\omega|\kappa|^2\frac{v_2
 T}{|x|^2}\log\frac{|x|}L\bigg)e^{-\omega|\kappa|^2\frac{v_2T}{|x|^2}\log
 \frac{|x|}L}\nn\\
&=\underbrace{\frac{O(1)}{\log\frac{|x|}L}\bigg(\frac{|x|^2}{v_2T\log\frac{|x|}L}
 \bigg)^{(d-2)/2}}_{(\log\frac{|x|}L)^{-1+(d-2)\mu/2}}\int_0^{\omega v_2TR^2\log
 \frac{|x|}L}\frac{\text{d}s}s~\bigg|\log\frac{s|x|^2}{\omega v_2T\log\frac{|x|}
 L}\bigg|s^{(d-2)/2}(1+s)e^{-s}.
\end{align}
Using the triangle inequality
\begin{align}
\bigg|\log\frac{s|x|^2}{\omega v_2T\log\frac{|x|}L}\bigg|\stackrel{\refeq{Tdef}}
 \le\bigg|\log\frac{L^2s}{\omega v_2}\bigg|+\mu\log\log\frac{|x|}L,
\end{align}
we obtain
\begin{align}
&\int_0^{\omega v_2TR^2\log\frac{|x|}L}\frac{\text{d}s}s~\bigg|\log\frac{s|x|^2}
 {\omega v_2T\log\frac{|x|}L}\bigg|s^{(d-2)/2}(1+s)e^{-s}\nn\\
&\le\underbrace{\int_0^\infty\frac{\text{d}s}s~\bigg|\log\frac{L^2s}{\omega
 v_2}\bigg|s^{(d-2)/2}(1+s)e^{-s}}_\text{convergent as long as $d>2$}+\big(
 \Gamma(\tfrac{d-2}2)+\Gamma(\tfrac{d}2)\big)\mu\log\log\frac{|x|}L\nn\\
&=O(1)\bigg(1+\log\log\frac{|x|}L\bigg).
\end{align}
As a result,
\begin{align}\lbeq{M3bd}
|M_3|\le\frac{O(1)\log\log\frac{|x|}L}{(\log\frac{|x|}L)^{1-(d-2)\mu/2}}
 \stackrel{\mu<\frac2{d+2}}\le\frac{O(1)\log\log\frac{|x|}L}{(\log\frac{|x|}
 L)^{4/(d+2)}},
\end{align}
which gives another error term.

Summarizing \refeq{S1dec2rewr}--\refeq{S1dec3} and \refeq{S1dec4}, we arrive at
\begin{align}
S_1(x)=\frac{|x|^{2-d}}{v_2\log\frac{|x|}L}\bigg(\gamma-\sum_{j=1}^3M_j\bigg)
 +\sum_{j=1}^4I_j,
\end{align}
with the error estimates \refeq{I1ubd}, \refeq{I2ubd}, 
\refeq{I3ubd}--\refeq{I4ubd}, \refeq{M1bd}, \refeq{M2bd} and \refeq{M3bd}.  
This completes the proof of Theorem~\ref{theorem:S} assuming the properties 
in Assumptions~\ref{assumption:hatD}--\ref{assumption:D}.
\QED

\subsection{Proof of the bound \refeq{Dnsupbd} on $\|D^{*n}\|_\infty$
for $\alpha=2$}\label{ss:D*n}
For $n=1$, 
$\|D\|_\infty=O(L^{-d})$ is obvious.  For $n\ge2$, we recall that
$\|D^{*n}\|_\infty$ is bounded as (cf., \cite[(A.2) and (A.4)]{csI})
\begin{align}\lbeq{D*n-unibd-pr}
\|D^{*n}\|_\infty\le O(L^{-d})\int_0^1\frac{\text{d}r}r~r^de^{-nr^2\log
 \frac\pi{2r}}+\|D\|_\infty(1-\Delta)^{n-2}.
\end{align}
Since the second term decays exponentially in $n$, it suffices to show that
\begin{align}
\int_0^1\frac{\text{d}r}r~r^de^{-nr^2\log\frac\pi{2r}}\le O\big((n\log
 \tfrac{\pi n}2)^{-d/2}\big).
\end{align}
Let $t=n^{-1/4}$ (so that $nt^2=\sqrt n$).  Notice that 
$\log\frac\pi{2r}\ge\log\frac\pi2>0$ for $r\le1$.  By changing variables as 
$s=nr^2\log\frac\pi2$, we have
\begin{align}
\int_t^1\frac{\text{d}r}r~r^de^{-nr^2\log\frac\pi{2r}}
&\le O(n^{-d/2})\int_{\sqrt n\log\frac\pi2}^\infty\frac{\text{d}s}s~s^{d/2}
 e^{-s}\le O(n^{-\frac{d+2}4})\,e^{-\sqrt n\log\frac\pi2},
\end{align}
which decays much faster than $O((n\log\frac{\pi n}2)^{-d/2})$.  For the 
remaining integral over $r\in(0,t)$, we change variables 
as $s=nr^2\log\frac\pi{2r}$.  Then, there is a $c>0$ such that
\begin{gather}\lbeq{changeofvariables}
\frac{\text{d}s}s=\bigg(2-\frac1{\log\frac\pi{2r}}\bigg)\frac{\text{d}r}r\ge
 \bigg(2-\frac1{\log\frac\pi{2t}}\bigg)\frac{\text{d}r}r\ge c\frac{\text{d}
 r}r,\\[5pt]
r=\sqrt{\frac{s}{n\log\frac\pi{2r}}}\le\sqrt{\frac{s}{n\log\frac\pi{2t}}}
 =\sqrt{\frac{s}{n(\frac14\log n+\log\frac\pi2)}}\le\sqrt{\frac{4s}{n\log\frac
 {\pi n}2}}.
\end{gather}
Therefore,
\begin{align}
\int_0^t\frac{\text{d}r}r~r^de^{-nr^2\log\frac\pi{2r}}\le\frac{4^{d/2}}{c
 (n\log\frac{\pi n}2)^{d/2}}\int_0^{nt^2\log\frac\pi{2t}}
 \frac{\text{d}s}s~s^{d/2}e^{-s}\le\frac{4^{d/2}\Gamma(\frac{d}2)}
 {c(n\log\frac{\pi n}2)^{d/2}},
\end{align}
as required.
\QED

\section{Analysis for the two-point function}\label{s:2pt}
In this section, we use the lace expansion \refeq{lace-intro} to prove 
Theorem~\ref{theorem:main}.  First, in Section~\ref{ss:basic}, we summarize 
some known facts, including the precise statement of the lace expansion for the 
two-point function.  Then, in Section~\ref{ss:xIRbd}, we prove the infrared 
bound \refeq{IRbd} by using convolution bounds on power functions with log 
corrections (Lemma~\ref{lemma:conv-bds}) and bounds on the lace-expansion 
coefficients (Lemma~\ref{lemma:diagbds}).  The proofs of those two lemmas 
follow, in Sections~\ref{ss:convbds}--\ref{ss:diagbds}, respectively.  Finally, in 
Section~\ref{ss:asymptotics}, we prove the asymptotic behavior \refeq{main} 
and complete the proof of Theorem~\ref{theorem:main}.

\subsection{List of known facts}\label{ss:basic}
The following four propositions hold independently of the value of $\alpha>0$.

\begin{proposition}[Lemma~2.2 of \cite{csIV}]\label{proposition:Gcont}
For every $x\in\Zd$, $G_p(x)$ is nondecreasing and continuous in $p<\pc$ for 
SAW, and in $p\le\pc$ for percolation and the Ising model.  The continuity up 
to $p=\pc$ for SAW is also valid if $G_p(x)$ is uniformly bounded in $p<\pc$.
\end{proposition}

\begin{proposition}[Lemma~2.3 of \cite{csIV}]\label{proposition:RWbds}
For every $p<\pc$ and $x\in\Zd$,
\begin{align}\lbeq{RWbds}
G_p(x)&\le S_p(x),&&
pD(x)\le G_p(x)-\delta_{o,x}\le(pD*G_p)(x).
\end{align}
\end{proposition}

\begin{proposition}[Lemma~2.4 of \cite{csIV}]\label{proposition:subpcbd}
For every $p<\pc$, there is a $K_p=K_p(\alpha,d,L)<\infty$ such that,
for any $x\in\Zd$,
\begin{align}\lbeq{subpcbd}
G_p(x)\le K_p\veee{x}_L^{-d-\alpha}.
\end{align}
\end{proposition}

\begin{proposition}[\cite{bs85} for SAW; \cite{hs90p} for percolation; 
\cite{s07} for the Ising model]\label{proposition:lace-exp}
There are model-dependent nonnegative functions on $\Zd$, 
$\{\pi_p^{\sss(n)}\}_{n=0}^\infty$ ($\pi_p^{\sss(0)}\equiv0$ for SAW) and 
$\{R_p^{\sss(n)}\}_{n=1}^\infty$, such that, for every integer $n\ge0$,
\begin{align}\lbeq{lace-precise}
G_p=
 \begin{cases}
 \delta+(pD_{\ne}+\pi_p^{\sss(\le n)})*G_p+(-1)^{n+1}R_p^{\sss(n+1)}
  &[\text{SAW}],\\[5pt]
 \pi_p^{\sss(\le n)}+\pi_p^{\sss(\le n)}*pD_{\ne}*G_p+(-1)^{n+1}R_p^{\sss(n+1)}
  \quad&[\text{percolation \& Ising}],
 \end{cases}
\end{align}
where the spatial variables are omitted (e.g., $G_p$ for $G_p(x)$, 
$\delta$ for $\delta_{o,x}$) and\footnote{The recursion equation 
\cite[(1.11)]{csIV} is correct for percolation and the Ising model, 
but not quite for SAW, as long as $D(o)>0$.  To deal with such $D$, 
the definition \cite[(1.13)]{csIV} of $\varPi_p$ needs slight modification.  
See \refeq{Pidef-SAW} below.}
\begin{align}\lbeq{pilen}
D_{\ne}=D-D(o)\delta,&&
\pi_p^{\sss(\le n)}=\sum_{j=0}^n(-1)^j\pi_p^{\sss(j)}.
\end{align}
Moreover, the remainder term obeys the following bound:
\begin{align}\lbeq{errorbds}
R_p^{\sss(n+1)}\le
 \begin{cases}
 \pi_p^{\sss(n+1)}*G_p&[\text{SAW}],\\[5pt]
 \pi_p^{\sss(n)}*pD*G_p\quad&[\text{percolation \& Ising}].
 \end{cases}
\end{align}
\end{proposition}

Before proceeding to the next subsection, we derive the unified expression 
\refeq{lace-intro} from \refeq{lace-precise}.  To do so, we first assume 
$p<\pc$ and $\sum_j\|\pi_p^{\sss(j)}\|_1<\infty$, which has been verified for 
$\alpha\ne2$, $d>\dc$ and $L\gg1$ in \cite{csIV} and is verified in the next 
subsection for $\alpha=2$, $d\ge\dc$ and $L\gg1$.  Then, by \refeq{errorbds}, 
we can take the $n\to\infty$ limit to obtain
\begin{align}\lbeq{lace-precise-limit}
G_p=
 \begin{cases}
 \delta+(pD_{\ne}+\pi_p)*G_p\quad&[\text{SAW}],\\
 \pi_p+\pi_p*pD_{\ne}*G_p&[\text{percolation \& Ising}],
 \end{cases}
\end{align}
where $\pi_p=\lim_{n\to\infty}\pi_p^{\sss(\le n)}$.  For percolation and the 
Ising model, if $pD(o)\|\pi_p\|_1<1$ (also verified for $\alpha\ne2$, $d>\dc$ 
and $L\gg1$ in \cite{csIV}, and for $\alpha=2$, $d\ge\dc$ and $L\gg1$ in the 
next subsection), then
\begin{eqnarray}
G_p&=&\pi_p+\pi_p*pD*G_p-pD(o)\pi_p*\underbrace{G_p}_\text{replace}\nn\\
&=&\pi_p+\pi_p*pD*G_p-pD(o)\pi_p*\Big(\pi_p+\pi_p*pD*G_p-pD(o)\pi_p*G_p\Big)
 \nn\\
&=&\big(\pi_p-pD(o)\pi_p^{*2}\big)+\big(\pi_p-pD(o)\pi_p^{*2}\big)*pD*G_p+\big(
 -pD(o)\big)^2\pi_p^{*2}*\underbrace{G_p}_\text{replace}\nn\\
&=&\big(\pi_p-pD(o)\pi_p^{*2}\big)+\big(\pi_p-pD(o)\pi_p^{*2}\big)*pD*G_p\nn\\
&&+\big(-pD(o)\big)^2\pi_p^{*2}*\Big(\pi_p+\pi_p*pD*G_p-pD(o)\pi_p*G_p\Big)\nn\\
&\vdots&\nn\\
&=&\varPi_p+\varPi_p*pD*G_p,
\end{eqnarray}
where
\begin{align}\lbeq{Pidef-perc&Ising}
\varPi_p=\pi_p+\sum_{n=1}^\infty\big(-pD(o)\big)^n\pi_p^{*(n+1)}.
\end{align}
For SAW, if $pD(o)+\|\pi_p\|_1<1$ (also verified for $\alpha\ne2$, $d>\dc$ and 
$L\gg1$ in \cite{csIV}, and for $\alpha=2$, $d\ge\dc$ and $L\gg1$ in the next 
subsection), then
\begin{eqnarray}
G_p&=&\delta+pD*G_p+\big(-pD(o)\delta+\pi_p\big)*\underbrace{G_p}_\text{replace}
 \nn\\
&=&\delta+pD*G_p+\big(-pD(o)\delta+\pi_p\big)*\Big(\delta+pD*G_p+\big(-pD(o)
 \delta+\pi_p\big)*G_p\Big)\nn\\
&=&\Big(\delta+\big(-pD(o)\delta+\pi_p\big)\Big)+\Big(\delta+\big(-pD(o)\delta
 +\pi_p\big)\Big)*pD*G_p\nn\\
&&+\big(-pD(o)\delta+\pi_p\big)^{*2}*\underbrace{G_p}_\text{replace}\nn\\
&\vdots&\nn\\
&=&\varPi_p+\varPi_p*pD*G_p,
\end{eqnarray}
where
\begin{align}\lbeq{Pidef-SAW}
\varPi_p=\delta+\sum_{n=1}^\infty\big(-pD(o)\delta+\pi_p\big)^{*n}.
\end{align}

\subsection{Proof of the infrared bound \refeq{IRbd}}\label{ss:xIRbd}
Let $\alpha=2$, $d\ge\dc$ and 
\begin{align}\lbeq{lambda}
\lambda=\sup_{x\ne o}\frac{S_1(x)}{\veee{x}_L^{2-d}/\log\veee{\frac{x}L}_1}
 =O(L^{-2}).
\end{align}
Define
\begin{align}\lbeq{g-def}
g_p=p\vee\sup_{x\ne o}\frac{G_p(x)}{\lambda\veee{x}_L^{2-d}/\log\veee{\frac{x}
 L}_1}.
\end{align}
We will show that $g_p$ satisfies the following three properties:
\begin{enumerate}[(i)]
\item
$g_p$ is continuous (and nondecreasing) in $p\in[1,\pc)$.
\item
$g_1\le1$.
\item
If $\lambda\ll1$ (i.e., $L\gg1$), then $g_p\le3$ implies $g_p\le2$ for every
$p\in(1,\pc)$.
\end{enumerate}
Notice that the above properties readily imply 
$G_p(x)\le2\lambda\veee{x}_L^{2-d}/\log\veee{\frac{x}L}_1$ for all $x\ne o$ and 
$p<\pc~(\le2)$.  By Proposition~\ref{proposition:Gcont}, we can extend this 
bound up to $\pc$, which completes the proof of \refeq{IRbd}.

It remains to prove the properties (i)--(iii).

\paragraph{\it Proof of (i).}
It suffices to show that
$\sup_{x\ne o}G_p(x)/\veee{x}_L^{2-d}/\log\veee{\frac{x}L}_1$ is continuous in
$p\in[1,p_0]$ for every fixed $p_0\in(1,\pc)$.  First, by the monotonicity of 
$G_p(x)$ in $p\le p_0$ and using Proposition~\ref{proposition:subpcbd}, we have
\begin{align}
\frac{G_p(x)}{\veee{x}_L^{2-d}/\log\veee{\frac{x}L}_1}\le\frac{G_{p_0}(x)}
 {\veee{x}_L^{2-d}/\log\veee{\frac{x}L}_1}\le\frac{K_{p_0}\veee{x}_L^{-d-2}}
 {\veee{x}_L^{2-d}/\log\veee{\frac{x}L}_1}=\frac{K_{p_0}}{\veee{x}_L^4/\log
 \veee{\frac{x}L}_1}.
\end{align}
On the other hand, for any $x_0\ne o$ with $D(x_0)>0$, there is an
$R=R(p_0,x_0)<\infty$ such that, for all $|x|\ge R$,
\begin{align}
\frac{K_{p_0}}{\veee{x}_L^4/\log\veee{\frac{x}L}_1}\le\frac{D(x_0)}
 {\veee{x_0}_L^{2-d}/\log\veee{\frac{x_0}L}_1}.
\end{align}
Moreover, by using $p\ge1$ and the lower bound of the second inequality
in \refeq{RWbds}, we have
\begin{align}
D(x_0)\le pD(x_0)\le G_p(x_0).
\end{align}
As a result, for any $p\in[1,p_0]$, we obtain
\begin{align}
\sup_{x\ne o}\frac{G_p(x)}{\veee{x}_L^{2-d}/\log\veee{\frac{x}L}_1}
 =\frac{G_p(x_0)}{\veee{x_0}_L^{2-d}\log\veee{\frac{x_0}L}_1}\vee\max_{x:
 0<|x|<R}\frac{G_p(x)}{\veee{x}_L^{2-d}/\log\veee{\frac{x}L}_1}.
\end{align}
Since $G_p(x)$ is continuous in $p$ (cf., Proposition~\ref{proposition:Gcont}) and the
maximum of finitely many continuous functions is continuous, we can conclude
that $g_p$ is continuous in $p\in[1,p_0]$, as required.
\QED

\paragraph{\it Proof of (ii).}
By Proposition~\ref{proposition:RWbds} and the definition \refeq{lambda} of $\lambda$, we readily 
obtain
\begin{align}
g_1=1\vee\sup_{x\ne o}\frac{G_1(x)}{\lambda\veee{x}_L^{2-d}/\log\veee{\frac{x}
 L}_1}\le1\vee\sup_{x\ne o}\frac{S_1(x)}{\lambda\veee{x}_L^{2-d}/\log
 \veee{\frac{x}L}_1}\le1,
\end{align}
as required.
\QED

\paragraph{\it Proof of (iii).}
This is the most involved part among (i)--(iii), and here we use the lace 
expansion.  To evaluate the lace-expansion coefficients, we use the following 
bounds on convolutions of power functions with log corrections, whose proof is 
deferred to Section~\ref{lemma:conv-bds}.

\begin{lemma}\label{lemma:conv-bds}
For $a_1\ge b_1>0$ with $a_1+b_1\ge d$, and for $a_2,b_2\ge0$ with
$a_2\ge b_2$ when $a_1=b_1$, there is an $L$-independent constant
$C=C(d,a_1,a_2,b_1,b_2)<\infty$ such that
\begin{align}\lbeq{conv1}
&\sum_{y\in\Zd}\frac{\veee{x-y}_L^{-a_1}}{(\log\veee{\frac{x-y}L}_1)^{a_2}}
 \frac{\veee{y}_L^{-b_1}}{(\log\veee{\frac{y}L}_1)^{b_2}}\\
&\le\frac{C\,\veee{x}_L^{-b_1}}{(\log\veee{\frac{x}L}_1)^{b_2}}\times
 \begin{cases}
 L^{d-a_1}&[a_1>d],\\
 \log\log\veee{\frac{x}L}_1&[a_1=d,~a_2=1],\\
 (\log\veee{\frac{x}L}_1)^{0\vee(1-a_2)}&[a_1=d,~a_2\ne1],\\
 \veee{x}_L^{d-a_1}&[a_1<d,~a_1+b_1>d],\\
 \veee{x}_L^{b_1}(\log\veee{\frac{x}L}_1)^{0\vee(1-a_2)}&[a_1<d,~a_1+b_1=d,
  ~a_2+b_2>1].
 \end{cases}\nn
\end{align}
\end{lemma}

Assuming $g_p\le3$ and Lemma~\ref{lemma:conv-bds},  we prove in 
Section~\ref{ss:diagbds} the following bounds on the lace-expansion 
coefficients $\{\pi_p^{\sss(n)}\}_{n=0}^\infty$ (recall that 
$\pi_p^{\sss(0)}\equiv0$ for SAW) in \refeq{lace-precise}.

\begin{figure}
\begin{align*}
{}_{\raisebox{-2pc}{$o$}}\!\!\!\raisebox{-3pc}{\includegraphics[scale=.4]
 {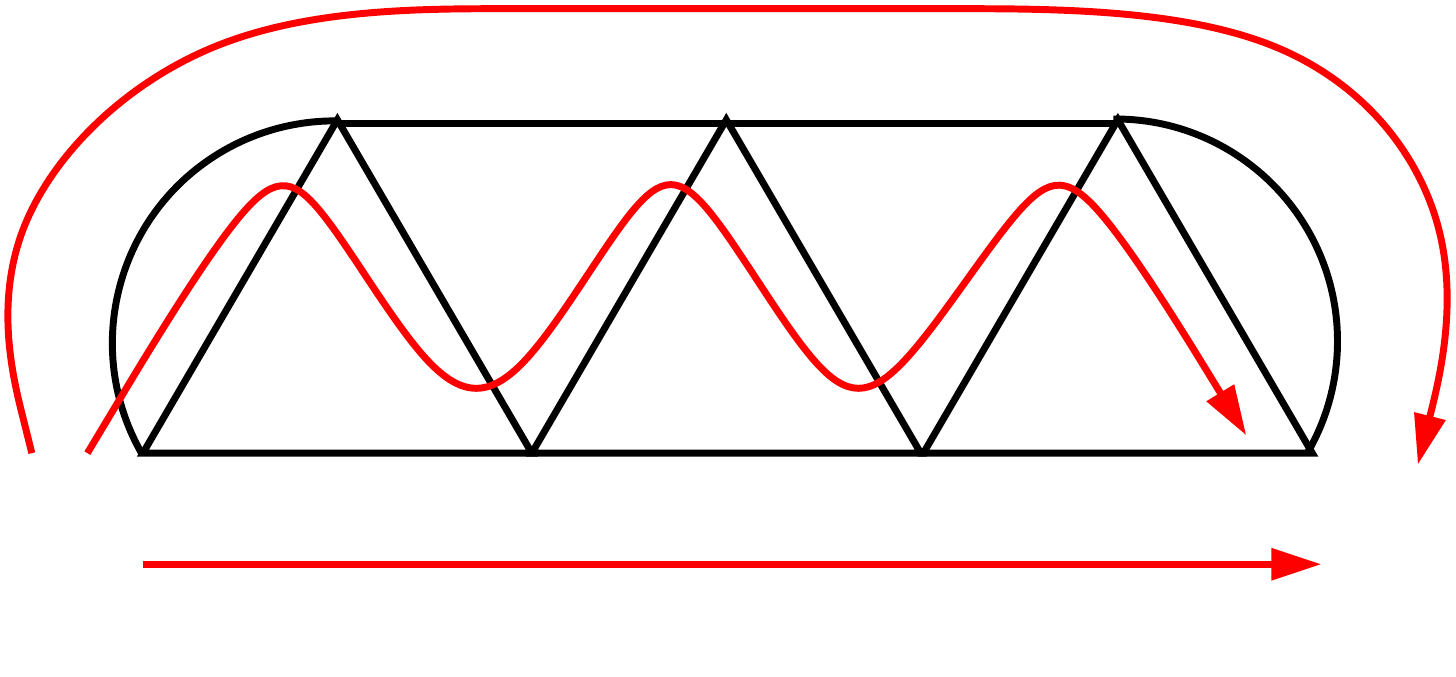}}\!\!\!{}_{\raisebox{-2pc}{$x$}}
&&&&
o\raisebox{-2.5pc}{\includegraphics[scale=.4]{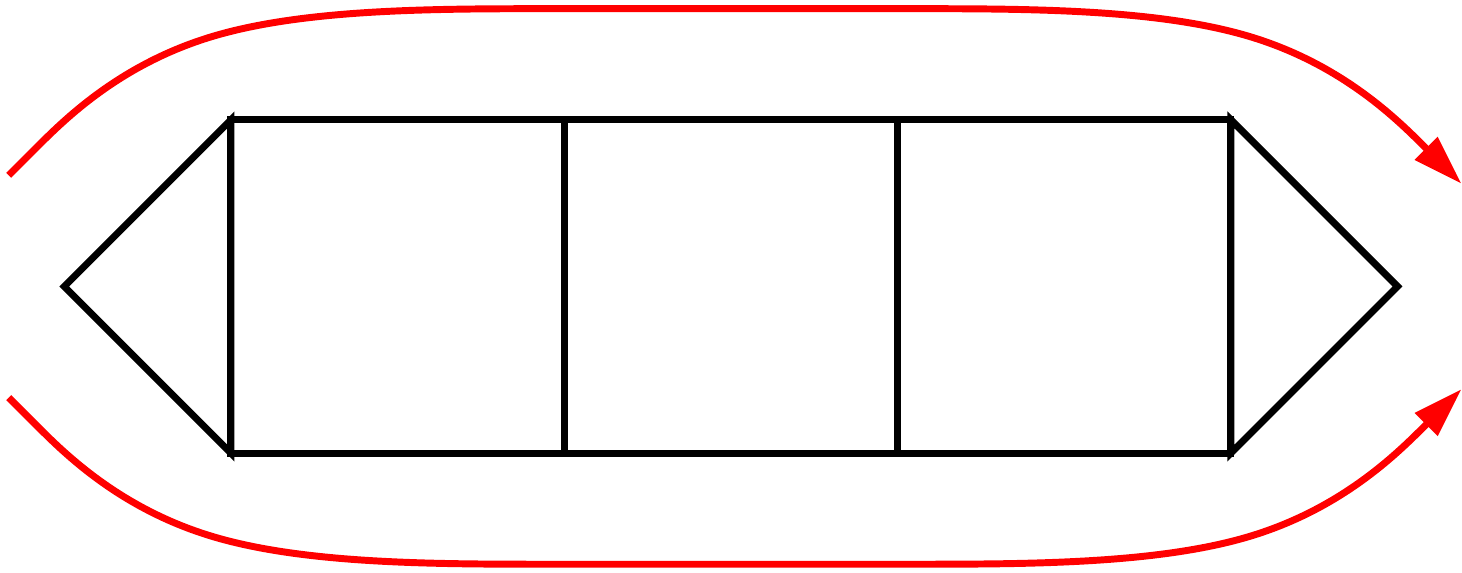}}x
\end{align*}
\caption{\label{fig:ell}
Examples of the lace-expansion diagrams for SAW and the Ising model (left) and 
percolation (right).  The factor $\ell$ in \refeq{elldef} is the number of 
disjoint paths (in red) from $o$ to $x$ using different sets of line segments.}
\end{figure}

\begin{lemma}\label{lemma:diagbds}
Let (cf., \refeq{dcdef} for the definition of $m$)
\begin{align}\lbeq{elldef}
\ell=\frac{m+1}{m-1}=
 \begin{cases}
 3\quad&[\text{SAW \& Ising}],\\
 2&[\text{percolation}].
 \end{cases}
\end{align}
Suppose $g_p\le3$ and $p<\pc$.  
Under the same condition as in Theorem~\ref{theorem:main}, 
we have
\begin{align}\lbeq{D*G-bd}
(pD*G_p)(x)\le O(\lambda)\frac{\veee{x}_L^{2-d}}{\log\veee{\frac{x}L}_1}
 \qquad[x\in\Zd].
\end{align}
Moreover, for SAW,
\begin{align}\lbeq{pijbds-SAW}
\pi_p^{\sss(j)}(x)\le
 \begin{cases}
 O(L^{-d})\delta_{o,x}&[j=1],\\[5pt]
 \dpst O(\lambda)^{j+1}\frac{\veee{x}_L^{3(2-d)}}{(\log\veee{\frac{x}
  L}_1)^3}\quad&[j\ge2],
 \end{cases}
\end{align}
and for the Ising model and percolation,
\begin{align}\lbeq{pijbds-Ising&perc}
\pi_p^{\sss(j)}(x)\le
 \begin{cases}
 \dpst O(L^{-d})^j\delta_{o,x}+O(\lambda)^2\frac{\veee{x}_L^{\ell(2-d)}}
  {(\log\veee{\frac{x}L}_1)^\ell}\quad&[j=0,1],\\[1pc]
 \dpst O(\lambda)^j\frac{\veee{x}_L^{\ell(2-d)}}{(\log\veee{\frac{x}L}_1)^\ell}
  &[j\ge2].
 \end{cases}
\end{align}
\end{lemma}

Consequently, we have $\sum_j\|\pi_p^{\sss(j)}\|_1<\infty$ for $d\ge\dc$ and 
$L\gg1$.  Then, by using \refeq{errorbds} for $p<\pc$, we obtain 
$\lim_{n\to\infty}\|R_p^{\sss(n)}\|_1=0$ and \refeq{lace-precise-limit} with
\begin{align}\lbeq{smallpibds}
|\pi_p(x)|\le
 \begin{cases}
 \dpst O(L^{-d})\delta_{o,x}+O(\lambda^3)\frac{\veee{x}_L^{3(2-d)}}{(\log
  \veee{\frac{x}L}_1)^3}&[\text{SAW}],\\[1pc]
 \dpst\big(1+O(L^{-d})\big)\delta_{o,x}+O(\lambda^2)\frac{\veee{x}_L^{\ell
  (2-d)}}{(\log\veee{\frac{x}L}_1)^\ell}&[\text{Ising \& percolation}].
 \end{cases}
\end{align}
This implies that, for SAW (cf., \refeq{Pidef-SAW}),
\begin{align}\lbeq{Pidef-SAW-factor}
\underbrace{pD(o)}_{O(L^{-d})}\delta_{o,x}+|\pi_p(x)|\le O(L^{-d})\delta_{o,x}
 +O(\lambda^3)\frac{\veee{x}_L^{3(2-d)}}{(\log\veee{\frac{x}L}_1)^3},
\end{align}
hence $pD(o)+\|\pi_p\|_1<1$ for $d\ge\dc$ and $L\gg1$.  Also, for the Ising 
model and percolation (cf., \refeq{Pidef-perc&Ising}), since 
$L^{-d}=O(\lambda)$ for $d\ge2$,
\begin{align}\lbeq{Pidef-perc&Ising-factor}
pD(o)|\pi_p(x)|\le O(L^{-d})\delta_{o,x}+O(\lambda^3)\frac{\veee{x}_L^{\ell(2
 -d)}}{(\log\veee{\frac{x}L}_1)^\ell},
\end{align}
hence $pD(o)\|\pi_p\|_1<1$ for $d\ge\dc$ and $L\gg1$.  We note that, 
for all three models,
\begin{align}\lbeq{ell2-d}
\ell(2-d)=-d-2-(\ell-1)(d-\dc).
\end{align}
By repeated applications of \refeq{Pidef-SAW-factor} and 
Lemma~\ref{lemma:conv-bds}, $\varPi_p(x)$ for SAW obeys the bound
\begin{align}\lbeq{varPi-bd-SAW}
&|\varPi_p(x)-\delta_{o,x}|\le\sum_{n=1}^\infty\big(pD(o)\delta+|\pi_p|\big)^{*
 n}(x)\nn\\
&\le\underbrace{\sum_{n=1}^\infty O(L^{-d})^n}_{O(L^{-d})}\delta_{o,x}
 +\sum_{n=1}^\infty\sum_{j=1}^n\binom{n}jO(L^{-d})^{n-j}O(\lambda^3)^j
 \underbrace{\frac{O(L^{-2-2(d-4)})^{j-1}\veee{x}_L^{3(2-d)}}{(\log
 \veee{\frac{x}L}_1)^3}}_{\because\,\text{Lemma~\ref{lemma:conv-bds}}}\nn\\
&\le O(L^{-d})\delta_{o,x}+O(\lambda^3)\frac{\veee{x}_L^{3(2-d)}}{(\log
 \veee{\frac{x}L}_1)^3}\underbrace{\sum_{n=1}^\infty n\Big(O(L^{-d})
 +\underbrace{O(\lambda^3L^{-2-2(d-4)})}_{O(L^{-2d})}\Big)^{n-1}}_{O(1)}.
\end{align}
Similarly, by repeated applications of \refeq{smallpibds}, 
\refeq{Pidef-perc&Ising-factor} and Lemma~\ref{lemma:conv-bds}, 
$\varPi_p(x)$ for the Ising model and percolation obeys the bound
\begin{align}\lbeq{varPi-bd-perc&Ising}
|\varPi_p(x)-\delta_{o,x}|&\le|\pi_p(x)-\delta_{o,x}|+\bigg(|\pi_p|*
 \underbrace{\sum_{n=1}^\infty\big(pD(o)|\pi_p|\big)^{*n}}_{\le
 \text{ RHS of \refeq{varPi-bd-SAW}}}\bigg)(x)\nn\\
&\le O(L^{-d})\delta_{o,x}+O(\lambda^2)\frac{\veee{x}_L^{3(2-d)}}{(\log
 \veee{\frac{x}L}_1)^3}.
\end{align}
By weakening the $O(\lambda^3)$ term in the right-most expression of 
\refeq{varPi-bd-SAW} to $O(\lambda^2)$, $\varPi_p(x)$ for all three models 
enjoys the unified bound
\begin{align}\lbeq{varPi-bd}
|\varPi_p(x)-\delta_{o,x}|&\le O(L^{-d})\delta_{o,x}+O(\lambda^2)\frac{
 \veee{x}_L^{\ell(2-d)}}{(\log\veee{\frac{x}L}_1)^\ell}.
\end{align}
As a result,
\begin{align}\lbeq{hatPi(0)bd}
|\hat\varPi_p(0)-1|\le O(L^{-d})+O(\lambda^2)\underbrace{\sum_x
 \frac{\veee{x}_L^{\ell(2-d)}}{(\log\veee{\frac{x}L}_1)^\ell}}_{O(L^{-2-(\ell-1)
 (d-\dc)})}=O(L^{-d}),
\end{align}
and
\begin{align}\lbeq{hatPidiffbd}
|\hat\varPi_p(0)-\hat\varPi_p(k)|\le O(\lambda^2)|k|^2\underbrace{\sum_x|x|^2
 \frac{\veee{x}_L^{\ell(2-d)}}{(\log\veee{\frac{x}L}_1)^\ell}}_{O(L^{-(\ell-1)
 (d-\dc)})}\le O(\lambda^2)|k|^2.
\end{align}

Now we are back to the proof of (iii).  First, by summing both sides of 
\refeq{lace-intro} over $x$ and solve the resulting equation for $\chi_p$, 
we have\begin{align}\lbeq{lace-again}
\chi_p=\hat\varPi_p(0)+\hat\varPi_p(0)p\chi_p=\frac{\hat\varPi_p(0)}{1-p\hat
 \varPi_p(0)}.
\end{align}
Since $\chi_p<\infty$ (because $p<\pc$) and $\hat\varPi_p(0)=1+O(L^{-d})>0$ 
for large $L$, we obtain
\begin{align}
p\hat\varPi_p(0)\in(0,1),
\end{align}
which implies $p<\hat\varPi_p(0)^{-1}=1+O(L^{-d})\le2$, as required.

Next, we investigate $G_p(x)$.  By repeated applications of \refeq{lace-intro} 
for $N$ times, we have
\begin{eqnarray}
G_p(x)&=&\varPi_p(x)+(\varPi_p*pD*G_p)(x)\nn\\
&=&\varPi_p(x)+(\varPi_p*pD*\varPi_p)(x)+\big((\varPi_p*pD)^{*2}*G_p\big)(x)
 \nn\\
&\vdots&\nn\\
&=&\bigg(\varPi_p*\sum_{n=0}^{N-1}(pD*\varPi_p)^{*n}\bigg)(x)+\big((\varPi_p*
 pD)^{*N}*G_p\big)(x).
\end{eqnarray}
Notice that, by \refeq{ell2-d}, \refeq{varPi-bd} and 
Lemma~\ref{lemma:conv-bds}, there are finite constants $C,C',C''$ such that
\begin{align}\lbeq{+ity}
(\varPi_p*D)(x)&\ge(1-CL^{-d})D(x)-C'\lambda^2\sum_y\frac{\veee{y}_L^{\ell(2
 -d)}}{(\log\veee{\frac{y}L}_1)^\ell}D(x-y)\nn\\
&\ge(1-CL^{-d}-C''\lambda^3)D(x),
\end{align}
which is positive for all $x$, if $L$ is large enough (see 
Remark~\ref{remark:new1step} below).  Therefore, 
\begin{align}\lbeq{new1step}
\cD(x)=\frac{(\varPi_p*D)(x)}{\hat\varPi_p(0)}
\end{align}
is a probability distribution that satisfies 
Assumptions~\ref{assumption:hatD}--\ref{assumption:D} (see computations 
below).  By this observation, we can take the limit
\begin{align}
0\le\big((\varPi_p*pD)^{*N}*G_p\big)(x)=\big(\underbrace{p\hat\varPi_p(0)}_{\in
 (0,1)}\big)^N\underbrace{(\cD^{*N}*G_p)(x)}_{\le\chi_p}\xrightarrow[N\to\infty]{}0,
\end{align}
so that
\begin{align}\lbeq{newID}
G_p(x)=\bigg(\varPi_p*\sum_{n=0}^\infty\big(p\hat\varPi_p(0)\big)^n\cD^{*n}
 \bigg)(x)=(\varPi_p*\cS_{p\hat\varPi_p(0)})(x),
\end{align}
where $\cS_q$ is the random-walk Green function generated by the 1-step 
distribution $\cD$ with fugacity $q\in[0,1]$, for which \refeq{Squbd} holds.  
By \refeq{varPi-bd} and Lemma~\ref{lemma:conv-bds}, we obtain that, 
for $x\ne o$,
\begin{align}
G_p(x)&\le\big(1+O(L^{-d})\big)\cS_1(x)+\underbrace{\sum_{y(\ne o)}\frac{O
 (\lambda^2)\veee{y}_L^{\ell(2-d)}}{(\log\veee{\frac{y}L}_1)^\ell}\bigg(
 \delta_{y,x}+\frac{O(\lambda)\veee{x-y}_L^{2-d}}{\log\veee{\frac{x-y}L}_1}
 \bigg)}_{O(\lambda^4)\veee{x}_L^{2-d}/\log\veee{\frac{x}L}_1}.
\end{align}
Suppose $\cS_1(x)\le(1+O(\lambda^3))S_1(x)$ holds for all $x$.  
Then, for $x\ne o$,
\begin{align}
G_p(x)&\le\big(1+O(\lambda^3)\big)S_1(x)+O(\lambda^4)\frac{\veee{x}_L^{2-d}}
 {\log\veee{\frac{x}L}_1}~\stackrel{L\gg1}\le~2\lambda\frac{\veee{x}_L^{2-d}}
 {\log\veee{\frac{x}L}_1},
\end{align}
as required.

It remains to show $\cS_1(x)\le(1+O(\lambda^3))S_1(x)$ for all $x$.  This is 
not so hard to verify, as explained now.  First, by \refeq{+ity} and its opposite 
inequality with all negative signs replaced by positive signs, 
\begin{align}
\bigg|\frac{\cD(x)}{D(x)}-1\bigg|=O(\lambda^3).
\end{align}
Also, by \refeq{hatPi(0)bd}--\refeq{hatPidiffbd} and \refeq{1-hatDbd2},
\begin{align}
\frac{1-\hat\cD(k)}{1-\hat D(k)}\stackrel{\refeq{new1step}}
 =1+\underbrace{\frac{\hat D(k)}{\hat\varPi_p(0)}}_{1+O(L^{-d})}
 \underbrace{\frac{\hat\varPi_p(0)-\hat\varPi_p(k)}{1-\hat D(k)}}_{O(\lambda^3)}
 =1+O(\lambda^3).
\end{align}
Similarly, 
\begin{align}\lbeq{v2again}
\frac{1-\hat\cD(k)}{|k|^2\log(1/|k|)}=\underbrace{\frac{1-\hat D(k)}{|k|^2\log(1
 /|k|)}}_{\to v_2,~\because\,\refeq{vdef}}+\underbrace{\frac{\hat D(k)}{\hat
 \varPi_p(0)}\frac{\hat\varPi_p(0)-\hat\varPi_p(k)}{|k|^2\log(1/|k|)}}_{\to0,~
 \because\,\refeq{hatPidiffbd}}~\xrightarrow[|k|\to0]{}v_2.
\end{align}
Therefore, for $L$ large enough, $\cD$ satisfies all 
\refeq{1-hatDbd1}--\refeq{Dnsupbd} with the same constants as $D$ (modulo 
$O(\lambda^3)$ terms).  Similar analysis can be applied to show that $\cD$ 
also satisfies \refeq{Dnxbd} with the same constant as $D$.  As a result, 
we can get $\cS_1(x)\le(1+O(\lambda^3))S_1(x)$ for all $x$.  
This completes the proof of (iii), hence the proof of the infrared bound 
\refeq{IRbd}.
\QED

\begin{remark}\label{remark:new1step}
{\rm
The above proof works as long as $\alpha\le2+(\ell-1)(d-\dc)$ (cf., 
\refeq{+ity}), then we can define the probability distribution \refeq{new1step} 
by taking $L$ sufficiently large. 
For short-range models investigated in \cite{h08,hhs03,s07}, on the other hand, 
since $\alpha$ is regarded as an arbitrarily large number, there is no way for 
\refeq{+ity} to be nonnegative for every $x$.  In this case, we may have to 
introduce a quite delicate function $E_{p,q,r}(x)$ as in \cite{csIV,hhs03} that 
is required to satisfy some symmetry conditions.  Since we do not need such 
a function for all $\alpha\le2$ and $d\ge\dc$, the analysis explained in this 
subsection is much easier and more transparent than the previous one in 
\cite{csIV,hhs03}. This is also related to the reason why the multiplicative 
constant $A$ in the asymptotic expression \refeq{previous} becomes 1 for 
$\alpha\le2$.
}
\end{remark}

\subsection{Convolution bounds on power functions with log corrections}
 \label{ss:convbds}
In this subsection, we prove Lemma~\ref{lemma:conv-bds}.  First, we rewrite 
the sum in \refeq{conv1} as
\begin{align}
\sum_{y\in\Zd}\frac{\veee{x-y}_L^{-a_1}}{(\log\veee{\frac{x-y}L}_1)^{a_2}}
 \frac{\veee{y}_L^{-b_1}}{(\log\veee{\frac{y}L}_1)^{b_2}}
&=\sum_{y:|x-y|\le|y|}\frac{\veee{x-y}_L^{-a_1}}{(\log\veee{\frac{x-y}
 L}_1)^{a_2}}\frac{\veee{y}_L^{-b_1}}{(\log\veee{\frac{y}L}_1)^{b_2}}\nn\\
&\quad+\sum_{y:|x-y|>|y|}\frac{\veee{x-y}_L^{-a_1}}{(\log\veee{\frac{x-y}
 L}_1)^{a_2}}\frac{\veee{y}_L^{-b_1}}{(\log\veee{\frac{y}L}_1)^{b_2}}\nn\\
&=\sum_{y:|x-y|\le|y|}\bigg(\frac{\veee{x-y}_L^{-a_1}}{(\log\veee{\frac{x-y}
 L}_1)^{a_2}}\frac{\veee{y}_L^{-b_1}}{(\log\veee{\frac{y}L}_1)^{b_2}}\nn\\
&\hskip5pc+\frac{\veee{x-y}_L^{-b_1}}{(\log\veee{\frac{x-y}L}_1)^{b_2}}
 \frac{\veee{y}_L^{-a_1}}{(\log\veee{\frac{y}L}_1)^{a_2}}\bigg).
\end{align}
Notice that the ratio of the second term to the first term in the parentheses, 
which is
\begin{align}
\frac{\dpst\frac{\veee{x-y}_L^{-b_1}}{(\log\veee{\frac{x-y}L}_1)^{b_2}}
 \frac{\veee{y}_L^{-a_1}}{(\log\veee{\frac{y}L}_1)^{a_2}}}{\dpst\frac{\veee{x
 -y}_L^{-a_1}}{(\log\veee{\frac{x-y}L}_1)^{a_2}}\frac{\veee{y}_L^{-b_1}}{(\log
 \veee{\frac{y}L}_1)^{b_2}}}
 =\bigg(\frac{\veee{x-y}_L}{\veee{y}_L}\bigg)^{a_1-b_1}\bigg(
 \frac{\log\veee{\frac{x-y}L}_1}{\log\veee{\frac{y}L}_1}\bigg)^{a_2-b_2},
\end{align}
is bounded above by an $L$-independent constant $C\in[1,\infty)$ as long as 
$a_1>b_1$, or $a_1=b_1$ and $a_2\ge b_2$.  Therefore,
\begin{align}\lbeq{conv1-pr1}
\sum_{y\in\Zd}\frac{\veee{x-y}_L^{-a_1}}{(\log\veee{\frac{x-y}L}_1)^{a_2}}
 \frac{\veee{y}_L^{-b_1}}{(\log\veee{\frac{y}L}_1)^{b_2}}
 \le2C\sum_{y:|x-y|\le|y|}\frac{\veee{x-y}_L^{-a_1}}{(\log\veee{\frac{x-y}
 L}_1)^{a_2}}\frac{\veee{y}_L^{-b_1}}{(\log\veee{\frac{y}L}_1)^{b_2}}.
\end{align}

Now we consider the following cases separately: (a) $a_1>d$, (b) $a_1=d$, 
(c) $a_1<d$ and $a_1+b_1\ge d$.
\begin{enumerate}[(a)]
\item
Let $a_1>d$.  Since $|x-y|\le|y|$ implies $|y|\ge\frac12|x|$, and since
\begin{align}\lbeq{veee-lbd}
\veee{\tfrac{x}2}_L\ge\frac12\veee{x}_L,&&
\log\veee{\tfrac{x}{2L}}_1\ge\frac{\log\frac\pi2}{\log\pi}\log\veee{\tfrac{x}
 L}_1,
\end{align}
we obtain
\begin{align}
\sum_{y:|x-y|\le|y|}\frac{\veee{x-y}_L^{-a_1}}{(\log\veee{\frac{x-y}L}_1)^{a_2}}
 \frac{\veee{y}_L^{-b_1}}{(\log\veee{\frac{y}L}_1)^{b_2}}\le\frac{O(1)
 \veee{x}_L^{-b_1}}{(\log\veee{\frac{x}L}_1)^{b_2}}\underbrace{\sum_{y\in\Zd}
 \frac{\veee{x-y}_L^{-a_1}}{(\log\veee{\frac{x-y}L}_1)^{a_2}}}_{O(L^{d-a_1})}.
\end{align}
\item
Let $a_1=d$.  First we split the sum as 
\begin{align}\lbeq{conv1-pr2}
\sum_{y:|x-y|\le|y|}=\sum_{\substack{y:|x-y|\le|y|\\ (|y|\le\frac32|x|)}}
 +\sum_{\substack{y:|x-y|\le|y|\\ (|y|>\frac32|x|)}}.
\end{align}
For the first sum, since $|x-y|\le|y|$ implies $|y|\ge\frac12|x|$ (so that 
\refeq{veee-lbd} holds), and since
\begin{align}\lbeq{veee-loglog}
\log\veee{\tfrac{3x}{2L}}_1\le\frac{\log\frac{3\pi}4}{\log\frac\pi2}\log
 \veee{\tfrac{x}L}_1,\hskip3pc
\log\log\veee{\tfrac{3x}{2L}}_1\le\frac{\log\log\frac{3\pi}4}{\log\log
 \frac\pi2}\log\log\veee{\tfrac{x}L}_1,
\end{align}
we obtain
\begin{align}\lbeq{conv1-pr3}
\sum_{\substack{y:|x-y|\le|y|\\ (|y|\le\frac32|x|)}}\frac{\veee{x-y}_L^{-d}}
 {(\log\veee{\frac{x-y}L}_1)^{a_2}}\frac{\veee{y}_L^{-b_1}}{(\log\veee{\frac{y}
 L}_1)^{b_2}}&\le\frac{O(1)\veee{x}_L^{-b_1}}{(\log\veee{\frac{x}L}_1)^{b_2}}
 \sum_{y:|x-y|\le\frac32|x|}\frac{\veee{x-y}_L^{-d}}{(\log\veee{\frac{x-y}
 L}_1)^{a_2}}\nn\\
\le\frac{O(1)\veee{x}_L^{-b_1}}{(\log\veee{\frac{x}L}_1)^{b_2}}&\times
 \begin{cases}
 1&[a_2>1],\\
 \log\log\veee{\frac{x}L}_1&[a_2=1],\\
 (\log\veee{\frac{x}L}_1)^{1-a_2}&[a_2<1].
 \end{cases}
\end{align}
For the second sum in \refeq{conv1-pr2}, since $|y|>\frac32|x|$ implies 
$|x-y|\ge\frac13|y|$, and since
\begin{align}\lbeq{veee-lbd2}
\veee{\tfrac{y}3}_L\ge\frac13\veee{y}_L,&&
\log\veee{\tfrac{y}{3L}}_1\ge\frac{\log\frac\pi2}{\log\frac{3\pi}2}\log
 \veee{\tfrac{y}L}_1,
\end{align}
we obtain
\begin{align}
\sum_{\substack{y:|x-y|\le|y|\\ (|y|>\frac32|x|)}}\frac{\veee{x
 -y}_L^{-d}}{(\log\veee{\frac{x-y}L}_1)^{a_2}}\frac{\veee{y}_L^{-b_1}}{(\log
 \veee{\frac{y}L}_1)^{b_2}}\le\frac{O(1)}{(\log\veee{\frac{x}L}_1)^{a_2+b_2}}
 \underbrace{\sum_{y:|y|>\frac32|x|}\veee{y}_L^{-d-b_1}}_{O(1)\veee{x}_L^{-
 b_1}},
\end{align}
which is smaller than \refeq{conv1-pr3}.
\item
Let $a_1<d$ and $a_1+b_1\ge d$.  Similarly to the case (b), we split the sum as 
in \refeq{conv1-pr2} and evaluate each sum by using \refeq{veee-lbd} and 
\refeq{veee-lbd2}.  Then, by discarding the log-dumping term 
$(\log\veee{\frac{x-y}L}_1)^{-a_2}$, the first sum 
in \refeq{conv1-pr2} is bounded as
\begin{gather}\lbeq{conv1-pr4}
\sum_{\substack{y:|x-y|\le|y|\\ (|y|\le\frac32|x|)}}\frac{\veee{x-y}_L^{-a_1}}
 {(\log\veee{\frac{x-y}L}_1)^{a_2}}\frac{\veee{y}_L^{-b_1}}{(\log\veee{\frac{y}
 L}_1)^{b_2}}\le\frac{O(1)\veee{x}_L^{-b_1}}{(\log\veee{\frac{x}L}_1)^{b_2}}
 \underbrace{\sum_{y:|x-y|\le\frac32|x|}\veee{x-y}_L^{-a_1}}_{O(1)\veee{x}_L^{d
 -a_1}}
\end{gather}
while the second sum in \refeq{conv1-pr2} is bounded as 
\begin{gather}
\sum_{\substack{y:|x-y|\le|y|\\ (|y|>\frac32|x|)}}\frac{\veee{x-y}_L^{-a_1}}
 {(\log\veee{\frac{x-y}L}_1)^{a_2}}\frac{\veee{y}_L^{-b_1}}{(\log\veee{\frac{y}
 L}_1)^{b_2}}\stackrel{\because\,|x-y|\ge\frac13|y|}\le O(1)\sum_{y:|y|>\frac32
 |x|}\frac{\veee{y}_L^{-a_1-b_1}}{(\log\veee{\frac{y}L}_1)^{a_2+b_2}}\nn\\
\le\frac{O(1)\veee{x}_L^{d-a_1-b_1}}{(\log\veee{\frac{x}L}_1)^{a_2+b_2}}\times
 \begin{cases}
 1&[a_1+b_1>d],\\
 \log\veee{\frac{x}L}_1&[a_1+b_1=d,~a_2+b_2>1],
 \end{cases}
\end{gather}
which is smaller (resp., larger) than \refeq{conv1-pr4} if $a_2>1$ (resp., 
$a_2<1$).  This completes the proof of Lemma~\ref{lemma:conv-bds}.
\QED
\end{enumerate}

\subsection{Bounds on the lace-expansion coefficients}\label{ss:diagbds}
In this subsection, we prove Lemma~\ref{lemma:diagbds}.  Suppose that $g_p\le3$ 
and 
$p<\pc$.  Since $G_p(y)=\delta_{o,y}+G_p(y)\ind{y\ne o}$ for all three models, 
we have
\begin{align}\lbeq{D*Gbd1}
(D*G_p)(x)=D(x)+\sum_{y\ne o}D(x-y)\,G_p(y).
\end{align}
The first term is easy, because
\begin{align}
D(x)=\frac{O(L^2)}{\veee{x}_L^{d+2}}=O(\lambda)\frac{\log\veee{\frac{x}L}_1}
 {\veee{\frac{x}L}_1^4}\frac{\veee{x}_L^{2-d}}{\log\veee{\frac{x}L}_1}\le O(\lambda)
 \frac{\veee{x}_L^{2-d}}{\log\veee{\frac{x}L}_1}.
\end{align}
For the second term in \refeq{D*Gbd1}, we use $g_p\le3$ and 
Lemma~\ref{lemma:conv-bds} as 
\begin{align}
\sum_{y\ne o}D(x-y)\,G_p(y)\le\sum_{y\in\Zd}\frac{O(L^2)}{\veee{x-y}_L^{d
 +2}}\frac{3\lambda\veee{y}_L^{2-d}}{\log\veee{\frac{y}L}_1}\le O(\lambda)
 \frac{\veee{x}_L^{2-d}}{\log\veee{\frac{x}L}_1}.
\end{align}
This completes the proof of \refeq{D*G-bd}.

To prove \refeq{pijbds-SAW}--\refeq{pijbds-Ising&perc}, we repeatedly apply 
Lemma~\ref{lemma:conv-bds} to the diagrammatic bounds on $\pi_p(x)$ in 
\cite{hhs03,s07}.  For example, the lace-expansion diagram in 
Figure~\ref{fig:ell} for SAW and the Ising model can be bounded as follows.  
Suppose for now that each line segment, say, from $x$ to $y$, represents 
$3\lambda\veee{x-y}_L^{2-d}/\log\veee{\frac{x-y}L}_1$, i.e., the assumed 
bound on the nonzero two-point function.  Then, by using 
Lemma~\ref{lemma:conv-bds} (to perform the sum over $w$), we can show 
that, for $d\ge4$,
\begin{align}
\raisebox{-1.9pc}{\includegraphics[scale=0.3]{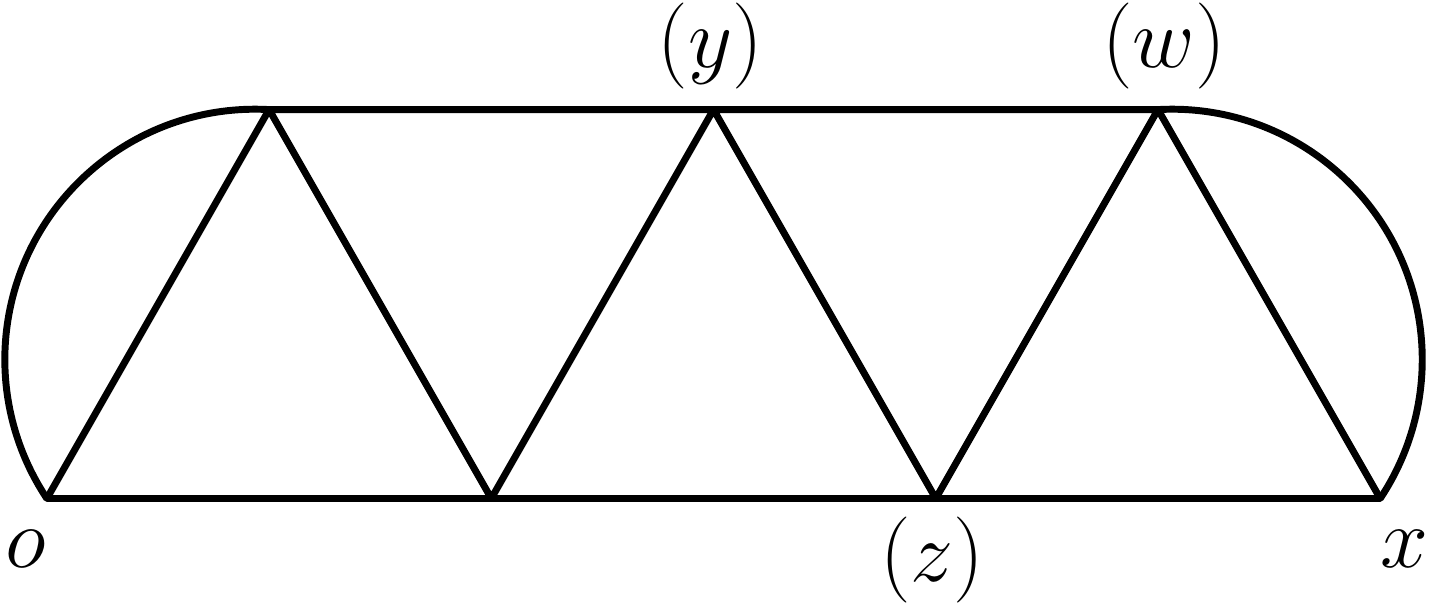}}~~\le4^{d-1}
 \bigg(3\lambda\frac{\log\pi}{\log\frac\pi2}\bigg)^2C~~\raisebox{-1.9pc}
 {\includegraphics[scale=0.3]{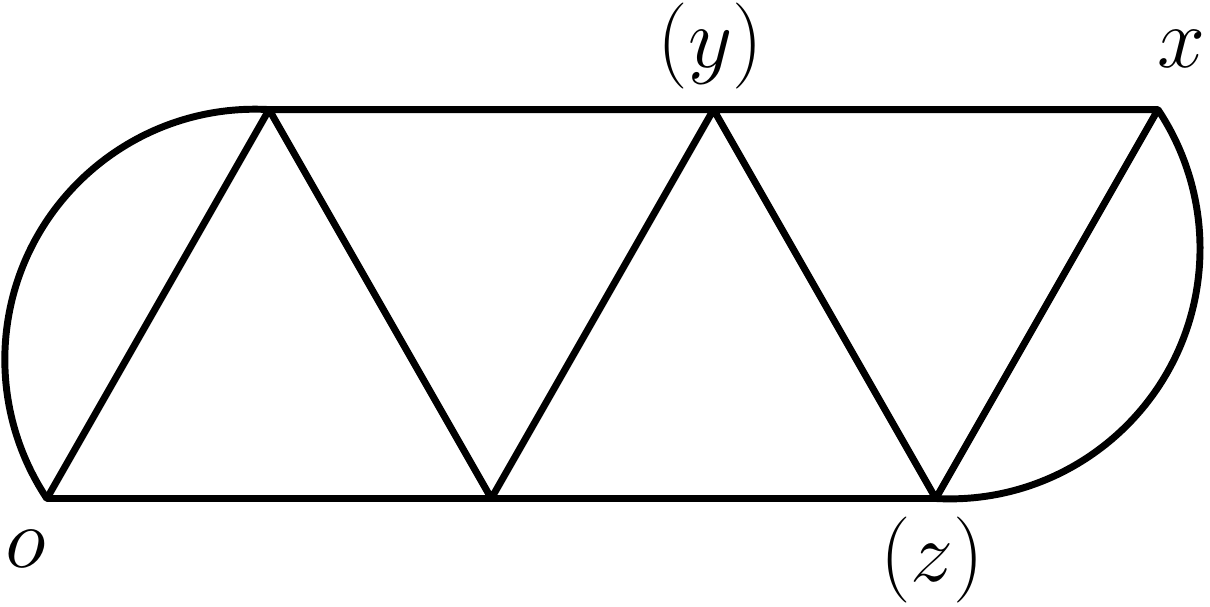}}~,
\end{align}
where the indicies in the parentheses are summed over $\Zd$.  This is due to 
the following computation: for $d\ge4$,
\begin{align}
&\sum_w\frac{3\lambda\veee{y-w}_L^{2-d}}{\log\veee{\frac{y-w}L}_1}\frac{3
 \lambda\veee{z-w}_L^{2-d}}{\log\veee{\frac{z-w}L}_1}\bigg(\frac{3\lambda
 \veee{w-x}_L^{2-d}}{\log\veee{\frac{w-x}L}_1}\bigg)^2\nn\\
&\qquad\times\Big(\ind{|y-w|\le|w-x|}+\ind{|y-w|\ge|w-x|}\Big)\Big(\ind{|z-w|
 \le|w-x|}+\ind{|z-w|\ge|w-x|}\Big)\nn\\
&\le\sum_w\frac{3\lambda\veee{y-w}_L^{2-d}}{\log\veee{\frac{y-w}L}_1}\frac{3
 \lambda\veee{z-w}_L^{2-d}}{\log\veee{\frac{z-w}L}_1}\bigg(\frac{3\lambda
 \veee{w-x}_L^{2-d}}{\log\veee{\frac{w-x}L}_1}\bigg)^2\underbrace{\ind{|y-w|\le
 |w-x|}}_{\Rightarrow|w-x|\ge\frac12|y-x|}~\underbrace{\ind{|z-w|\le|w-x|}}_{
 \Rightarrow|w-x|\ge\frac12|z-x|}\nn\\
&\quad+[\text{3 other cases}]\nn\\[5pt]
&\le(3\lambda)^2\frac{3\lambda\veee{\frac{y-x}2}_L^{2-d}}{\log\veee{\frac{y-x}
 {2L}}_1}\frac{3\lambda\veee{\frac{z-x}2}_L^{2-d}}{\log\veee{\frac{z-x}{2L}}_1}
 \bigg(\underbrace{\underbrace{\sum_w\frac{\veee{y-w}_L^{2-d}}{\log\veee{\frac{
 y-w}L}_1}\frac{\veee{z-w}_L^{2-d}}{\log\veee{\frac{z-w}L}_1}}_{\le C\veee{y-
 z}_L^{2-d}/\log\veee{\frac{y-z}L}_1}+[\text{3 other cases}]}_{\le4C}\bigg)\nn\\
&\le4^{d-1}\bigg(3\lambda\frac{\log\pi}{\log\frac\pi2}\bigg)^2
 C~\frac{3\lambda\veee{y-x}_L^{2-d}}{\log\veee{\frac{y-x}L}_1}\frac{3\lambda
 \veee{z-x}_L^{2-d}}{\log\veee{\frac{z-x}L}_1}.
\end{align}
By repeated application of the above inequality, we will end up with 
\begin{align}
\raisebox{-1.9pc}{\includegraphics[scale=0.3]{sawIsing1}}~~&\le\Bigg(4^{d-1}
 \bigg(3\lambda\frac{\log\pi}{\log\frac\pi2}\bigg)^2C\Bigg)^5~~\raisebox{-1.8pc}
 {\includegraphics[scale=0.3]{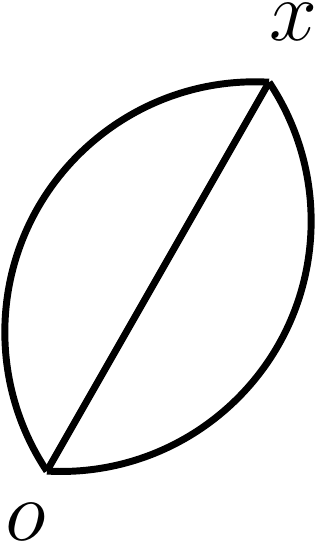}}\nn\\
&\le\underbrace{\Bigg(4^{d-1}\bigg(3\lambda\frac{\log\pi}{\log\frac\pi2}\bigg)^2
 C\Bigg)^5(3\lambda)^3}_{O(\lambda)^{13}}\frac{\veee{x}_L^{3(2-d)}}{(\log
 \veee{\frac{x}L}_1)^3},
\end{align}
which is smaller than \refeq{pijbds-SAW}--\refeq{pijbds-Ising&perc}, by a 
factor $O(\lambda)^5$ for SAW, in particular.  This is because, in fact, not 
every line segment is nonzero.  The situation for the Ising model and 
percolation is harder, because most of the line segments can be zero-length, 
which do not have small factors of $\lambda$.  However, the convolution 
$pD*G_p$ shows up repeatedly, which has a small factor of $\lambda$, as 
in \refeq{D*G-bd}.  This also provides a bound on the main contribution from 
$\pi_p^{\sss(1)}(x)$, as
\begin{align}
(pD*G_p)(o)\delta_{o,x}\le3\bigg(D(o)+\sum_{y\ne o}\frac{O(L^2)}{\veee{y}_L^{d
 +2}}\frac{3\lambda\veee{y}_L^{2-d}}{\log\veee{\frac{y}L}_1}\bigg)\delta_{o,x}
 =O(L^{-d})\delta_{o,x}.
\end{align}
This completes the sketch proof of Lemma~\ref{lemma:diagbds}.
\QED

\subsection{Proof of the asymptotic behavior \refeq{main}}
\label{ss:asymptotics}
First we recall \refeq{newID}.  Since 
$\chi_p=\hat\varPi_p(0)/(1-p\hat\varPi_p(0))$ diverges as $p\uparrow\pc$, 
while $\hat\varPi_p(0)=1+O(L^{-d})$ uniformly in $p<\pc$, we have 
$\pc\hat\varPi_{\pc}(0)=1$.  Therefore,
\begin{align}\lbeq{Gpcasy1}
G_{\pc}(x)=\varPi_{\pc}*\cS_1(x)=\underbrace{\hat\Pi_{\pc}(0)}_{1/\pc}\cS_1(x)
 +\sum_{y\ne o}\Pi_{\pc}(y)\big(\cS_1(x-y)-\cS_1(x)\big).
\end{align}
The asymptotic expression of $\cS_1(x)$ is the same as that of $S_1(x)$.  This 
can be shown by following the proof of \refeq{S1asy} and using the limit 
\refeq{v2again}.

To investigate the error term in \refeq{Gpcasy1}, we first split the sum as 
\begin{align}\lbeq{splitting}
\sum_{y\ne o}~=\sum_{y:0<|y|\le\frac13|x|}+\sum_{y:|x-y|\le\frac13|x|}
 +\sum_{y:|y|\wedge|x-y|>\frac13|x|}\equiv~{\sum_y}'+{\sum_y}''+{\sum_y}'''.
\end{align}
For $\sum_y''$, since $|x-y|\le\frac13|x|$ implies $\frac23|x|\le|y|$ (so that 
a similar inequality to \refeq{veee-lbd} or \refeq{veee-lbd2} holds), we 
have that, for large $|x|$,
\begin{eqnarray}\lbeq{f*g''}
\bigg|{\sum_y}''\varPi_{\pc}(y)\big(\cS_1(x-y)-\cS_1(x)\big)\bigg|
&\stackrel{\refeq{varPi-bd}}\le&\frac{O(\lambda^2)|x|^{\ell(2-d)}}{(\log
 |x|)^\ell}\underbrace{\sum_{y:|x-y|\le\frac13|x|}\big(\cS_1(x-y)+\cS_1(x)
 \big)}_{O(\lambda)|x|^2/\log|x|}\nn\\
&=&O(\lambda^3)\frac{|x|^{-d-(\ell-1)(d-\dc)}}{(\log|x|)^{\ell+1}}.
\end{eqnarray}
Similarly, for $\sum_y'''$ for large $|x|$,
\begin{eqnarray}\lbeq{f*g'''}
\bigg|{\sum_y}'''\varPi_{\pc}(y)\big(\cS_1(x-y)-\cS_1(x)\big)\bigg|
&\stackrel{\refeq{varPi-bd}}\le&\frac{O(\lambda)|x|^{2-d}}{\log|x|}
 \underbrace{\sum_{y:|y|>\frac13|x|}\frac{O(\lambda^2)|y|^{\ell(2-d)}}{(\log
 |y|)^\ell}}_{O(\lambda^2)|x|^{-2-(\ell-1)(d-\dc)}/(\log|x|)^\ell}\nn\\
&=&O(\lambda^3)\frac{|x|^{-d-(\ell-1)(d-\dc)}}{(\log|x|)^{\ell+1}}.
\end{eqnarray}

It remains to investigate $\sum_y'$.  For that, we first use \refeq{S1asy} and 
the $\Zd$-symmetry of $\varPi_{\pc}$ to obtain that, for large $|x|$,
\begin{align}\lbeq{sum'}
&\bigg|{\sum_y}'\varPi_{\pc}(y)\big(\cS_1(x-y)-\cS_1(x)\big)\bigg|\nn\\
&\le\underbrace{\frac{\gamma_2}{v_2}}_{O(\lambda)}{\sum_y}'|\varPi_{\pc}(y)|
 \bigg|\frac12\bigg(\frac{|x+y|^{2-d}}{\log|x+y|}+\frac{|x-y|^{2-d}}{\log|x-y|}
 \bigg)-\frac{|x|^{2-d}}{\log|x|}\bigg|\nn\\
&\quad+\underbrace{\sum_{y:0<|y|\le\frac13|x|}|\varPi_{\pc}(y)|}_{O(\lambda^3)}
 \frac{O(\lambda)|x|^{2-d}}{(\log|x|)^{1+\epsilon}}.
\end{align}
Then, by Taylor's theorem, 
\begin{align}
|x\pm y|^{2-d}&=|x|^{2-d}\bigg(1\pm2\frac{x\cdot y}{|x|^2}+\frac{O(|y|^2)}
 {|x|^2}\bigg)^{(2-d)/2}\nn\\
&=|x|^{2-d}\bigg(1\pm(2-d)\frac{x\cdot y}{|x|^2}+\frac{O(|y|^2)}{|x|^2}\bigg),\\
\log|x\pm y|&=\log|x|+\log\frac{|x\pm y|}{|x|}=\log|x|\pm\frac{x\cdot y}{|x|^2}
 +\frac{O(|y|^2)}{|x|^2},
\end{align}
which implies 
\begin{align}
\bigg|\frac12\bigg(\frac{|x+y|^{2-d}}{\log|x+y|}+\frac{|x-y|^{2-d}}{\log|x-y|}
 \bigg)-\frac{|x|^{2-d}}{\log|x|}\bigg|\le \frac{O(|y|^2)|x|^{-d}}{\log|x|}.
\end{align}
Therefore, the first term on the right-hand side of \refeq{sum'} is bounded by
\begin{align}\lbeq{f*g':|x|>2L}
\frac{O(\lambda)|x|^{-d}}{\log|x|}\underbrace{\sum_{y:0<|y|\le\frac13|x|}\frac{O
 (\lambda^2)\veee{y}_L^{-d-(\ell-1)(d-\dc)}}{(\log\veee{\frac{y}L}_1)^\ell}}_{O
 (\lambda^2)}=O(\lambda^3)\frac{|x|^{-d}}{\log|x|}.
\end{align}
This completes the proof of Theorem~\ref{theorem:main}.
\QED

\section*{Acknowledgements}
The work of AS was supported by JSPS KAKENHI Grant Number 18K03406.  
The work of LCC was supported by the grant MOST 107-2115-M-004 -004 -MY2.  
We are grateful to the Institute of Mathematics and Mathematics Research 
Promotion Center (MRPC) of Academia Sinica, as well as the National Center 
for Theoretical Sciences (NCTS) at National Taiwan University, for providing 
us with comfortable working environment in multiple occasions.  We would also 
like to thank the referees for their comments to improve presentation of this 
paper.

\end{document}